\documentclass{article}
\usepackage{multicol}
\usepackage{ragged2e}
\usepackage{graphicx,setspace}
\usepackage{natbib}
\usepackage{subfig}
\usepackage{fancyhdr,amsmath,amssymb, amsthm,mathtools}
\usepackage{etoolbox}
\usepackage[ruled,norelsize,vlined]{algorithm2e}

\usepackage[colorlinks,citecolor=blue,urlcolor=blue]{hyperref}
\usepackage{booktabs}
\usepackage{tikz}
\usetikzlibrary{matrix,positioning}
\usetikzlibrary{fit,positioning,calc}
\usepackage{scrextend}
\usepackage{hyperref}
\usepackage{chngcntr}
\usepackage{apptools}
\usepackage{bbm}
\usepackage{ulem}
\usepackage{caption}
\usepackage[T1]{fontenc}
\usepackage{tgtermes}
\usepackage[small]{titlesec}
\usepackage[export]{adjustbox}
\usepackage{multirow}
\usepackage{cuted}

\newcommand{\bb}{ {\bf b} }

\newcommand{\bx}{ {\bf x} }
\newcommand{\bP}{ {\bf P} }

\newcommand{\bs}{ {\bf s} }
\newcommand{\bU}{ {\bf U} }
\newcommand{\by}{ {\bf y} }
\newcommand{\bX}{ {\bf X} }

\newcommand{\bD}{ {\bf D} }
\newcommand{\bI}{ {\bf I} }
\newcommand{\bQ}{ {\bf Q} }
\newcommand{\bV}{ {\bf V} }

\newcommand{\bG}{ {\bf G} }
\newcommand{\bg}{ {\bf g} }
\newcommand{\bW}{ {\bf W} }
\newcommand{\bA}{ {\bf A} }
\newcommand{\bB}{ {\bf B} }

\newcommand{\btt}{ {\bf t} }
\newcommand{\bv}{ {\bf v} }
\newcommand{\ba}{ {\bf a} }
\newcommand{\br}{ {\bf r} }

\newcommand{\bgamma}{ {\boldsymbol \gamma} }

\newcommand{\bvarepsilon}{ {\boldsymbol \varepsilon} }
\newcommand{\bLambda}{ {\boldsymbol \Lambda} }
\newcommand{\bPsi}{ {\boldsymbol \Psi} }

\newcommand{\bGamma}{ {\boldsymbol \Gamma} }
\newcommand{\bDelta}{ {\boldsymbol \Delta} }
\newcommand{\bbeta}{ {\boldsymbol \beta} }
\newcommand{\bmu}{ {\boldsymbol \mu} }
\newcommand{\bSigma}{ {\boldsymbol \Sigma} }

\newcommand{\bOmega}{ {\boldsymbol \Omega} }
\newcommand{\bomega}{ {\boldsymbol \omega} }

\newcommand{\bxi}{ {\boldsymbol \xi} }
\newcommand{\bz}{ {\bf z} }

\newcommand{\boeta}{ {\boldsymbol \eta} }
\newcommand{\bpsi}{ {\boldsymbol \psi} }

\newcommand{\pst}{\mbox{\normalfont \tiny post}}
\newcommand{\argmin}{\mbox{argmin}} 
\newcommand{\argmax}{\mbox{argmax}}

\newtheorem{Theorem}{Theorem}

\newtheorem{Lemma}{Lemma}

\begin{document}

\title{Bayesian Conjugacy in Probit, Tobit, Multinomial Probit and Extensions: \\ A Review and New Results}
	\author{Niccol\`o Anceschi$^{\mbox{\small a}}$, Augusto Fasano$^{\mbox{\small b}}$, Daniele Durante$^{\mbox{\small a}}$\footnote{Department of Decision Sciences and Bocconi Institute for Data Science and Analytics, Bocconi University, Italy, e-mail: daniele.durante@unibocconi.it}, and Giacomo Zanella$^{\mbox{\small a}}$ \\ {\small $^{\mbox{\small a}}$Department of Decision Sciences and Bocconi Institute for Data Science and Analytics, Bocconi University, Milan, Italy}, \\ {\small $^{\mbox{\small b}}$Collegio Carlo Alberto, Turin, Italy}}
	\date{}
	\maketitle

\begin{abstract}
A broad class of models that routinely appear in several fields can be expressed as partially or fully discretized Gaussian linear regressions. Besides including classical Gaussian response settings, this class also encompasses probit, multinomial probit and tobit regression, among others, thereby yielding one of the most widely-implemented families of models in routine applications. The relevance of such representations has stimulated decades of  research in the Bayesian field, mostly motivated by the fact that, unlike for Gaussian linear regression, the posterior distribution induced by such models does not seem to belong to a known class, under the commonly-assumed Gaussian priors for the  coefficients. This has motivated several  solutions for posterior inference relying either on sampling-based strategies or on deterministic approximations that, however, still experience computational and accuracy  issues, especially in high dimensions. The scope of this article is to review, unify and extend recent advances in Bayesian inference and computation for this core class of models. To address such a goal, we prove that the likelihoods induced by these formulations share a common analytical structure implying conjugacy with a broad class of distributions, namely the unified skew-normal (SUN), that generalize Gaussians to include skewness. This result unifies and extends recent conjugacy properties  for specific models within the class analyzed, and  opens new avenues for improved posterior inference, under a broader class of formulations and priors, via novel closed-form expressions, i.i.d.\ samplers from the exact SUN posteriors, and more accurate and scalable approximations from variational Bayes and expectation-propagation. Such advantages are illustrated in  simulations  and are expected to facilitate the routine-use of these core Bayesian models, while providing novel frameworks for studying theoretical properties and developing future extensions.  
\end{abstract}
\noindent%
{\it Keywords:} Bayesian computation, Data augmentation, Expectation-propagation, Truncated normal distribution, Unified skew-normal distribution, Variational Bayes.

\begin{multicols}{2}

\section{Introduction}\label{sec_1}

The scope of this contribution is to review, unify, compare and extend both past and recent developments in Bayesian inference for  probit \citep{bliss1934method}, multinomial probit \citep{hausman1978,tutz1991,stern_1992} and tobit \citep{Tobin} regression models, along with related extensions to multivariate, skewed, non-linear  and dynamic contexts. Although such models are core formulations in statistics \citep[][]{demaris2004regression,Greene2003,agrest_2013} and often appear as building-blocks in more complex constructions \citep[see e.g.,][]{chipman_2010,rodrig_2011}, Bayesian inference under the associated likelihoods still presents open challenges that have motivated decades of active research in the field  \citep[][]{chopin_2017}. This is mainly due to the presence in the likelihood of Gaussian cumulative distribution functions arising from a partial or full discretization of a set of continuous latent utilities under a discrete choice perspective \citep[e.g.,][]{Greene2003}, thus  hindering Gaussian conjugacy when combined with~the commonly-assumed multivariate normal priors for the coefficients in $\bbeta$. 

This lack of conjugacy for such a  routinely-implemented class of models motivates ongoing efforts to develop effective MCMC-based sampling methods and accurate deterministic approximations of the posterior distribution to perform Bayesian inference in probit \citep{Albert_1993, holmes_2006,consonni_2007, chopin_2017},  tobit \citep{CHIB199279,chib2009_tob,loaiza2021}, multinomial probit \citep{Albert_1993,mcculloch1994,nobile1998,mcculloch2000,albert2001,imai2005,loaiza2021scalable}, and their  multivariate, skewed, dynamic  and non-linear generalizations \citep[][]{Chib1998,chen1999new,andrieu2002particle,sahu2003new,kuss2005assessing,girolami2006,bazan2010framework,talhouk2012efficient,soyer2013bayesian,riihimki2012nested}. Although these methods yield state-of-the-art implementations, there are still open questions on computational scalability, approximation accuracy and mixing, especially in high dimensions  \citep[][]{chopin_2017}. Such issues, combined with the recent conjugacy results for probit models in \citet{durante_2019}, have led to  renewed interest in closed-form solutions for Bayesian inference under these formulations. In particular,  \citet{durante_2019} proved that the posterior for the $\bbeta$ coefficients in Bayesian probit regression under Gaussian priors belongs to the class of unified skew-normal  (SUN) distributions \citep{ARELLANO_VALLE_2006} and, more generally, that SUNs are conjugate to probit regression. The SUN class extends~multivariate Gaussians to include skewness, and its analytical properties have led to rapid extensions of the original results in \citet{durante_2019} to multinomial probit \citep{fasano2020}, dynamic multivariate probit \citep{fasano2021closed}, Gaussian processes \citep{cao2020scalable}, skewed Gaussian processes \citep{benavoli2020skew,benavoli2021unified}, skew-elliptical link functions \citep{zhang2021tractable}  and rounded data \citep{kowal2021conjugate}, while facilitating the development of more accurate approximations \citep{fasano2019}. 

These advancements provide yet unexplored opportunities for Bayesian inference under such models via novel closed-form expressions, tractable Monte Carlo methods based on i.i.d.\ samples from the exact SUN posteriors, and more accurate and scalable approximations from variational Bayes (VB)  \citep[e.g.,][]{blei_2017} and expectation-propagation~(EP) \citep[e.g.,][]{chopin_2017}. However, most of these new developments focus on specific sub-classes of models within a potentially broader family of formulations that rely on partially or fully discretized Gaussian latent utilities. Therefore, there is still the lack of a unified framework which  would be practically and conceptually useful to derive general conjugacy results along with broadly-applicable closed-form solutions,  Monte Carlo methods and improved approximations of the posterior distribution. For instance, conjugacy results for tobit models \citep{Tobin} are yet missing in the literature, however, as it will be clarified in Section~\ref{sec_3}, SUNs are conjugate also to this class. Such a comprehensive treatment would also help to clarify these advancements in the light of previously-developed state-of-the-art MCMC methods and approximations, and would serve as a catalyst of applied, methodological and theoretical research to further expand the set of solutions for this broad class of models. 

This article aims at covering the aforementioned gap to boost the routine use of these core Bayesian models, and provide comprehensive frameworks for studying general theoretical properties and developing future extensions. As a first step toward accomplishing this goal, Section~\ref{sec_2} unifies probit, tobit, multinomial probit and related extensions  by reformulating the associated likelihoods as special cases of a  general form that relies on  the product between multivariate Gaussian densities and cumulative distributions, both evaluated at a linear combination of the coefficients $\bbeta$. Such a unified formulation is crucial to prove a general result in Section~\ref{sec_3} which states that SUN distributions are conjugate priors for any model whose likelihood can be expressed as a special case of the one defined in  Section~\ref{sec_2}. This result unifies available findings for probit \citep{durante_2019}, multinomial probit \citep{fasano2020} and dynamic multivariate probit \citep{fasano2021closed}, among others, while extending SUN conjugacy properties to a  broader class of Bayesian models for which similar results have not appeared yet in the literature. Notable examples are tobit \citep{Tobin}, and any extension of probit, tobit and multinomial probit  which replaces the Gaussian latent utilities with skew-normal ones \citep{chen1999new,sahu2003new,bazan2010framework}, among others; see Section~S1 in the Supplementary Materials. 

As discussed in Section~\ref{sec_4}, this unified conjugacy result is also practically relevant since it allows to inherit all the recent methodological and computational developments for Bayesian inference under SUN posteriors in probit and multinomial probit  to the entire class of models presented in Section~\ref{sec_2}. These advancements include novel closed-form expressions for relevant posterior moments, marginal likelihoods and predictive distributions, along with improved Monte Carlo methods and deterministic approximations from VB and EP. These solutions are presented in detail within Section~\ref{sec_4} along with a careful review of the previous state-of-the-art solutions recasted under the proposed general framework. An excellent overview of these previous strategies can be  found in  \citet{chopin_2017}, but the focus is on univariate probit models. Due to this, the present article will mostly consider the more recent developments relying on SUN conjugacy and on their discussion in the light of previous solutions, when adapted to the broader class of models and priors, beyond classical Bayesian probit regression. Consistent with this scope, Section~\ref{sec_6} concludes with a general discussion that points towards several future research directions motivated by the unified framework developed in this article. Empirical studies illustrating the potentials of this unification are provided in Section~\ref{sec_5}. The proofs of the theoretical results in this article and an in-depth discussion of the computational costs for the methods presented in Section~\ref{sec_4} can be found in the Supplementary Materials. \texttt{R} codes with tutorials to reproduce the analyses in Section~\ref{sec_5} are available at  \url{https://github.com/niccoloanceschi/TobitSUN}.

\section{A unified likelihood representation}\label{sec_2}

As discussed in Section~\ref{sec_1}, probit regression \citep{bliss1934method}, tobit  \citep{Tobin}, multinomial probit \citep{hausman1978,tutz1991,stern_1992} and their extensions are core formulations within statistics, and, when viewed as specific examples of a more general representation which also includes classical Gaussian linear regression, arguably yield to one of the most widely-implemented classes of models in routine applications \citep[][]{demaris2004regression,Greene2003,agrest_2013}. In fact, all these formulations share a common generative construction, in that the corresponding responses can be defined as partially or fully discretized versions of  continuous ones from a set of underlying  Gaussian linear regressions \citep{CHIB199279,Albert_1993,Chib1998}. In particular, let $z_i \in \mathbbm{R}$ denote a latent continuous response available for each unit $i=1, \ldots, n$, and consider the standard linear regression model $z_i= \bx^{\intercal}_i\bbeta+\varepsilon_i$, with i.i.d. noise $\varepsilon_i \sim \mbox{N}(0, \sigma^2)$, covariates' vector $\bx_i=(x_{i1}, \ldots, x_{ip})^{\intercal}$ and coefficients $\bbeta=(\beta_1, \ldots, \beta_p)^{\intercal}$. Starting from this building-block formulation, general Gaussian linear regression models, probit models \citep{bliss1934method} and tobit regression \citep{Tobin} can be obtained by letting  $y_i=z_i$, $y_i=\mathbbm{1}(z_i>0)$ and $y_i=\max\{z_i,0\}=z_i\mathbbm{1}(z_i>0)$, respectively. The first two constructions correspond to the limiting cases in which $z_i$ is either entirely observed or dichotomized, respectively, whereas the third one represents the intermediate situation in which $z_i$ is fully observed only if it exceeds value $0$  \citep{CHIB199279,Albert_1993}. Multinomial probit regression \citep{hausman1978,tutz1991,stern_1992} for categorical responses $y_i \in \{1, \ldots, L\}$ can be derived with a similar reasoning. For instance, in the formulation proposed by \citet{stern_1992}, the observed categorical response $y_i$ is defined as $y_i= \mbox{argmax}_l\{z_{i1}, \ldots, z_{iL}\}$   where $z_{i1}, \ldots, z_{iL}$ are class-specific Gaussian latent utilities related to the covariates via a system of linear regressions $z_{il}= \bx^{\intercal}_i\bbeta_l+\varepsilon_{il}$ for each $l=1, \ldots, L$, with $\bvarepsilon_i=(\varepsilon_{i1}, \ldots, \varepsilon_{iL})^{\intercal} \sim \mbox{N}_L({\bf 0}, \bSigma)$.

As shown  in Sections~\ref{sec_2.1}--\ref{sec_2.4}, these similarities in the generative models imply that  the likelihoods induced by the above formulations and their extensions are all specific examples of the general form
\begin{equation}
\begin{split}
p(\by  \mid \bbeta)&=p(\bar{\by}_1 \mid \bbeta)p(\bar{\by}_0  \mid \bbeta)\\
& \propto  \phi_{\bar{n}_1}(\bar{\by}_1-\bar{\bX}_1 \bbeta;\bar{\bSigma}_1)\Phi_{\bar{n}_0}(\bar{\by}_0+\bar{\bX}_0 \bbeta; \bar{\bSigma}_0),
\end{split}
\label{eq1}
\end{equation}
with $\phi_{\bar{n}_1}(\bar{\by}_1-\bar{\bX}_1 \bbeta; \bar{\bSigma}_1)$ and $\Phi_{\bar{n}_0}(\bar{\by}_0+\bar{\bX}_0 \bbeta; \bar{\bSigma}_0)$ denoting the density and the cumulative distribution function of the multivariate Gaussians $\mbox{N}_{\bar{n}_1}({\bf 0}, \bar{\bSigma}_1)$ and $\mbox{N}_{\bar{n}_0}({\bf 0}, \bar{\bSigma}_0)$, evaluated at $\bar{\by}_1-\bar{\bX}_1 \bbeta$ and $\bar{\by}_0+\bar{\bX}_0 \bbeta$, respectively, where $\bar{\by}_1:=\bar{\by}_1(\by)$ and $\bar{\by}_0:=\bar{\by}_0(\by)$ denote known response vectors obtained as a function of $\by$, whereas $\bar{\bX}_1:=\bar{\bX}_1(\by, \bX)$ and $\bar{\bX}_0:=\bar{\bX}_0(\by, \bX)$ are suitable matrices which can be directly derived from the original design matrix $\bX$ --- whose rows comprise the vectors $\bx^{\intercal}_i$, $i=1, \ldots, n$ --- and, possibly, the responses in $\by$. Intuitively, Equation~\eqref{eq1} states that the likelihood for $\by$ arises from a set of linear regressions for $\bar{n}_1+\bar{n}_0$ Gaussian responses of which $\bar{n}_1$ are fully observed, thus contributing to the first factor in~\eqref{eq1}, whereas the remaining $\bar{n}_0$ are dichotomized and hence are captured by the second. For instance, the likelihood $\prod_{i=1}^n\Phi(\bx^{\intercal}_i\bbeta)^{\mathbbm{1}(y_i=1)}[1-\Phi(\bx^{\intercal}_i\bbeta)]^{\mathbbm{1}(y_i=0)}$ under probit regression can be rewritten as $\prod_{i=1}^n \Phi[(2y_i-1)\bx_i^{\intercal}\bbeta]=\Phi_{n}(\mbox{diag}(2\by-{\bf 1}_n)\bX\bbeta; \bI_n)$, that coincides with  \eqref{eq1} after letting $\bar{n}_1=0$, $\bar{n}_0=n$, $\bar{\by}_0={\bf 0}$, $\bar{\bX}_0=\mbox{diag}(2\by-{\bf 1}_n)\bX$ and $\bar{\bSigma}_0=\bI_n$. This example is a degenerate case where $\bar{n}_1=0$, meaning that there are no fully-observed Gaussian responses and  $\bar{n}_0=n$ dichotomized ones; refer to Sections~\ref{sec_2.1}--\ref{sec_2.4}  for additional examples which further clarify Equation~\eqref{eq1}. To simplify notation, in the following we write $\phi\left(\cdot\right)$ and $\Phi\left(\cdot\right)$ for denoting, respectively, the density and cumulative distribution function of a univariate standard Gaussian variable, with mean $0$ and variance $1$.

\subsection{Linear regression and multivariate linear regression}\label{sec_2.1}

Although the focus of this article is on models beyond the classical Gaussian response setting, it is worth emphasizing that also this class induces likelihoods which are special cases of \eqref{eq1}. For instance,  Gaussian linear regression $(y_i \mid \bbeta) \sim \mbox{N}(\bx^{\intercal}_i\bbeta, \sigma^2)$, independently for $i=1, \ldots, n$, is directly recovered after noticing that the induced likelihood
\begin{equation}
\begin{split}
p(\by \mid \bbeta)&=  \prod\nolimits_{i=1}^{n} \phi(y_i- \bx^{\intercal}_i\bbeta; \sigma^2)\\
&= \phi_{n}(\by-\bX\bbeta; \sigma^2\bI_n),
\end{split}
\label{eq2}
\end{equation}
coincides with \eqref{eq1}, when letting $\bar{n}_0=0$, $\bar{n}_1=n$, $\bar{\by}_1=\by$, $\bar{\bX}_1=\bX$ and $\bar{\bSigma}_1=\sigma^2\bI_n$. As a consequence, also heteroscedastic and correlated versions can be incorporated by replacing $\sigma^2\bI_n$ with a general covariance matrix. Similarly, the likelihood associated with multivariate Gaussian response data from $(\by_i \mid \bbeta) \sim \mbox{N}_m(\bX_i\bbeta, \bSigma)$, independently for $i=1, \ldots, n$, can be written as
\begin{equation}
\begin{split}
p(\by \mid \bbeta)&= \prod\nolimits_{i=1}^{n} \phi_m(\by_i- \bX_i\bbeta; \bSigma)\\
&= \phi_{n \cdot m}(\by-\bX\bbeta; \bI_n \otimes \bSigma),
\end{split}
\label{eq3}
\end{equation}
where $\by=(\by^{\intercal}_1, \ldots, \by^{\intercal}_n)^{\intercal}$, $\bX=(\bX^{\intercal}_1, \ldots, \bX^{\intercal}_n)^{\intercal}$ and $\otimes$ denotes the Kronecker product. Setting $\bar{n}_0=0$, $\bar{n}_1=n \cdot m$, $\bar{\by}_1=\by$, $\bar{\bX}_1=\bX$ and $\bar{\bSigma}_1=\bI_n \otimes \bSigma$ in  \eqref{eq1} yields directly to \eqref{eq3}. Notice that, unlike for the probit example introduced in Section~\ref{sec_2}, in these cases $\bar{n}_0=0$, meaning that all the Gaussian responses are fully observed and there are no dichotomized outcomes. This translates into the fact that no contribution by the Gaussian cumulative distribution function is present in likelihoods \eqref{eq2}--\eqref{eq3}, which, therefore, simplify to multivariate Gaussian densities.

\subsection{Probit, multivariate probit and multinomial probit}\label{sec_2.2}

As discussed in Section~\ref{sec_2}, the classical probit regression model $(y_i \mid \bbeta) \sim \mbox{Bern}[\Phi(\bx^{\intercal}_i\bbeta)]$, independently for $i=1, \ldots, n$, induces likelihoods which can be readily reframed within representation \eqref{eq1}. More specifically, recalling \citet{durante_2019} and denoting with ${\bf 1}_n$ the $(n \times 1)$--dimensional vector of all ones,  the probit likelihood  can be expressed as
\begin{equation}
\begin{split}
p(\by \mid \bbeta)&=\prod\nolimits_{i=1}^n \Phi[(2y_i-1)\bx_i^{\intercal}\bbeta]\\
&=\Phi_{n}(\mbox{diag}(2\by-{\bf 1}_n)\bX\bbeta; \bI_n),
\end{split}
\label{eq5}
\end{equation}
which is a special case of \eqref{eq1}, with $\bar{n}_1=0$, $\bar{n}_0=n$, $\bar{\by}_0={\bf 0}$, $\bar{\bX}_0=\mbox{diag}(2\by-{\bf 1}_n)\bX$ and $\bar{\bSigma}_0=\bI_n$. Replacing $\bar{\bSigma}_0=\bI_n$ with $\bar{\bSigma}_0=\sigma^2\bI_n$ yields also probabilities of the form $\Phi(\bx^{\intercal}_i\bbeta; \sigma^2)$.

The above probit model further admits a number of routinely-used extensions which incorporate multivariate binary outcomes \citep{Chib1998} and also multinomial response data \citep{hausman1978,tutz1991,stern_1992}. As previously mentioned, both cases have their roots in discrete choice models \citep[e.g.,][]{Greene2003}, and can be reframed within \eqref{eq1}. To clarify this result, let us first focus on multivariate probit models for the binary response vector $\by_i=(y_{i1}, \ldots, y_{im})^{\intercal} \in \{0, 1\}^{m}$. As discussed in \citet{Chib1998}, 
these formulations can be interpreted as a dichotomized version of the  regression model for multivariate Gaussian response data  in Section~\ref{sec_2.1}. In fact, each  $\by_i$ is defined as $\by_i=[\mathbbm{1}(z_{i1}>0), \ldots, \mathbbm{1}(z_{im}>0)]^{\intercal}={\boldsymbol{\mathbbm{1}}}(\bz_i>{\bf 0})$, where $(\bz_i=(z_{i1}, \ldots, z_{im})^{\intercal} \mid \bbeta) \sim \mbox{N}_m(\bX_i\bbeta, \bSigma)$, independently for every $i=1, \ldots, n$. This means that the contribution to the likelihood of each unit $i$ is  $p(\by_i \mid \bbeta)=p({\boldsymbol{\mathbbm{1}}}(\bz_i>{\bf 0}) \mid \bbeta)$, which can be also written as $\Phi_m(\bB_i \bX_i \bbeta; \bB_i \bSigma \bB_i)$, following standard properties of multivariate Gaussian cumulative distribution functions, where $\bB_i= \mbox{diag}(2 y_{i1}-1, \ldots, 2 y_{im}-1)$. As a result, the joint likelihood of multivariate probit  is 
\begin{equation}
\begin{split}
p(\by \mid \bbeta)&= \prod\nolimits_{i=1}^{n}\Phi_m(\bB_i \bX_i \bbeta; \bB_i \bSigma \bB_i)\\
&=\Phi_{n \cdot m}(\bB \bX \bbeta; \bB(\bI_n \otimes \bSigma) \bB),
\end{split}
\label{eq6}
\end{equation}
where $\bX=(\bX^{\intercal}_1, \ldots, \bX^{\intercal}_n)^{\intercal}$, and $\bB$ denotes an $(n \cdot m) \times( n \cdot m)$ block-diagonal matrix with generic block $\bB_{[i,i]}=\bB_i$, for each $i=1, \ldots, n$. To reframe \eqref{eq6} within the general likelihood form in \eqref{eq1} it suffices to set $\bar{n}_1=0$, $\bar{n}_0=n \cdot m$, $\bar{\by}_0={\bf 0}$, $\bar{\bX}_0=\bB\bX$ and $\bar{\bSigma}_0=\bB(\bI_n \otimes \bSigma) \bB$. 

As discussed in \citet{fasano2020}, a similar construction and derivations can be also considered for several multinomial probit models  \citep{hausman1978,tutz1991,stern_1992}. All these formulations express the probabilities of the $L$ different categories $\{1, \ldots, L\}$ via a discrete choice mechanism relying on correlated predictor-dependent  Gaussian latent utilities which facilitate improved flexibility and avoid restrictive assumptions often found in multinomial logit, such as the independence of irrelevant alternatives \citep{hausman1978}. For instance, in the formulation by \citet{stern_1992}, each categorical response $y_i$ is defined as $y_i= \mbox{argmax}_l\{z_{i1}, \ldots, z_{iL}\}$, where  $z_{il}= \bx^{\intercal}_i\bbeta_l+\varepsilon_{il}$ for $l=1, \ldots, L$, with $\bvarepsilon_i=(\varepsilon_{i1}, \ldots, \varepsilon_{iL})^{\intercal} \sim \mbox{N}_L({\bf 0}, \bSigma)$, and $\bbeta_L={\bf 0}$ for identifiability purposes \citep{johndrow2013}. As a consequence of this construction and recalling  Section~2.2 in \citet{fasano2020}, it  follows that $\mbox{pr}(y_i=l \mid \bbeta)=\mbox{pr}(z_{il}> z_{ik}, \forall k \neq l)$, which can be also re-written as $\mbox{pr}(\varepsilon_{ik}-\varepsilon_{il}<\bx^{\intercal}_{i}\bbeta_l-\bx^{\intercal}_{i}\bbeta_k,  \forall k \neq l)$. Therefore, let $\bv_l$ denote the $L$--dimensional vector with value $1$ in position $l$ and $0$ elsewhere, for each $l=1, \ldots, L$, and define $\bx_{il}=\bar{\bv}_l \otimes \bx_i$, where $\bar{\bv}_l$ is the $(L-1)$--dimensional vector obtained by removing the $L$--th entry in $\bv_l$. Then, $\mbox{pr}(y_i=l \mid \bbeta)=\mbox{pr}[(\bv_k-\bv_l)^{\intercal}\bvarepsilon_i<(\bx_{il}-\bx_{ik})^{\intercal}\bbeta, \forall k \neq l]$, where $\bbeta=(\bbeta^{\intercal}_1, \ldots, \bbeta^{\intercal}_{L-1})^{\intercal}$. This expression can be also re-formulated in the more compact form $\mbox{pr}(\bV_{[-l]}\bvarepsilon_i<\bX_{i[-l]}\bbeta)$, where $\bV_{[-l]}$ and $\bX_{i[-l]}$ correspond to suitable matrices whose rows are obtained by stacking the vectors $(\bv_k-\bv_l)^{\intercal}$ and $(\bx_{il}-\bx_{ik})^{\intercal}$, respectively, for every $k\neq l$, and $<$ is intended elementwise. Therefore, leveraging the standard properties of multivariate Gaussians, it follows that $\mbox{pr}(y_i=l \mid \bbeta)=\mbox{pr}(\bV_{[-l]}\bvarepsilon_i<\bX_{i[-l]}\bbeta)= \Phi_{L-1}(\bX_{i[-l]}\bbeta; \bV_{[-l]} \bSigma \bV^{\intercal}_{[-l]})$, for every $l=1, \ldots, L$. This result yields  a joint likelihood for the categorical responses $\by=(y_1, \ldots, y_n)^{\intercal}$ which can be written as
\begin{equation}
\begin{split}
p(\by \mid \bbeta)&= \prod\nolimits_{i=1}^{n}\Phi_{L-1}(\bX_{i[-y_i]}\bbeta; \bV_{[-y_i]} \bSigma \bV^{\intercal}_{[-y_i]})\\
&=\Phi_{n \cdot (L-1)}(\bX \bbeta; \bV(\bI_n \otimes \bSigma) \bV^{\intercal}),
\end{split}
\label{eq7}
\end{equation}
where $\bX=(\bX^{\intercal}_{1[-y_1]}, \ldots, \bX^{\intercal}_{n[-y_n]})^{\intercal}$, and $\bV$ is a block-diagonal matrix with generic block $\bV_{[i,i]}=\bV_{[-y_i]}$, for each $i=1, \ldots, n$. Setting $\bar{n}_1=0$, $\bar{n}_0=n \cdot (L-1)$, $\bar{\by}_0={\bf 0}$, $\bar{\bX}_0=\bX$ and $\bar{\bSigma}_0=\bV(\bI_n \otimes \bSigma) \bV^{\intercal}$ in \eqref{eq1} leads to Equation \eqref{eq7}. Hence, the multinomial probit model by \citet{stern_1992} is again a special case of the general form in \eqref{eq1}.  As shown in Sections 2.1 and 2.3 of \citet{fasano2020}, also the alternative formulations proposed by \citet{hausman1978} and \citet{tutz1991} induce likelihoods which can be expressed as cumulative distribution functions of multivariate Gaussians evaluated at a suitable linear combination of the coefficients' vector $\bbeta$; see Propositions 1 and 3 in  \citet{fasano2020}. This means that also such models can be easily recasted within the general form in \eqref{eq1} with $\bar{n}_1=0$, $\bar{\by}_0={\bf 0}$, and suitably defined $\bar{n}_0$, $\bar{\bX}_0$ and $\bar{\bSigma}_0$.  

\subsection{Tobit regression}\label{sec_2.3}

Recalling Section~\ref{sec_2}, the classical tobit model  \citep{Tobin} characterizes the intermediate situation in which response data are fully observed only if exceeding a certain threshold, often set to  $0$. This means that $y_i=z_i\mathbbm{1}(z_i>0)$, with  $(z_i \mid \bbeta) \sim \mbox{N}(\bx^{\intercal}_i\bbeta, \sigma^2)$, independently for $i=1, \ldots, n$. Such a formulation yields the joint likelihood
	\begin{equation}
	\begin{split}
	p(\by \mid \bbeta)&= {\prod\nolimits_{i:y_i{>}0}} \phi(y_i{-} \bx^{\intercal}_i \bbeta{;} \sigma^2){\prod\nolimits_{i:y_i=0}}\Phi(-\bx^{\intercal}_i \bbeta{;} \sigma^2)\\
	&=  \phi_{n_1}(\by_1-\bX_1\bbeta; \sigma^2\bI_{n_1})\Phi_{n_0}(-\bX_{0} \bbeta; \sigma^2\bI_{n_0}),
		\end{split}
	\label{eq9}
	\end{equation}
where $n_1$ and $n_0$ denote the number of fully observed and censored units, respectively, whereas $\by_1$, $\bX_1$ and $\bX_0$ are the response vector and design matrices associated with these two subsets of units. This likelihood can be again expressed as a special example of Equation \eqref{eq1} by letting $\bar{n}_1=n_1$, $\bar{n}_0=n_0$, $\bar{\by}_1=\by_1$, $\bar{\by}_0={\bf 0}$, $\bar{\bX}_1=\bX_1$, $\bar{\bX}_0=-\bX_0$, $\bar{\bSigma}_1=\sigma^2\bI_{n_1}$ and $\bar{\bSigma}_0=\sigma^2\bI_{n_0}$. 

The above result also holds for several subsequent extensions of the original  tobit model  \citep{Tobin}, which include more elaborated censoring mechanisms, possibly relying on multivariate Gaussian utilities. Such generalizations, often known in the literature as type {\rm II}, {\rm III}, {\rm IV}, and {\rm V} tobit models, are carefully discussed in \citet{amemiya1984} and all induce likelihoods which can be written as the product of Gaussian densities and cumulative distribution functions evaluated at suitable linear combinations of the coefficients $\bbeta$. This common structure allows again to readily express such extensions as special cases of the general form in \eqref{eq1}. Inclusion of multivariate versions is also straightforward under a similar reasoning considered in \eqref{eq3} and \eqref{eq6}.

\subsection{Extensions to skewed, non-linear and dynamic models}\label{sec_2.4}

As discussed in Section~S1 of the Supplementary Materials, although the models discussed in Sections~\ref{sec_2.1}--\ref{sec_2.3} cover the most widely-implemented formulations in the literature, several additional extensions of these representations to skewed \citep[e.g.,][]{chen1999new,sahu2003new,bazan2010framework,hutton2011modelling}, non-linear  \citep[e.g.,][]{kuss2005assessing,de2005bayesian,nickisch2008approximations,riihimki2012nested,cao2020scalable,benavoli2020skew}, dynamic \citep[e.g.,][]{manrique1998simulation,andrieu2002particle,naveau2005skewed,chib2006inference,soyer2013bayesian,fasano2021closed} and other contexts admit  likelihood forms as in~\eqref{eq1}.

\section{Conjugacy via unified skew-normals}\label{sec_3}

Sections~\ref{sec_3.1}--\ref{sec_3.2} unify Bayesian inference for the whole family of models  in Section~\ref{sec_2} by proving that the  likelihood in \eqref{eq1} admits as conjugate priors the class of unified skew-normal (SUN) distributions \citep{ARELLANO_VALLE_2006}. Crucially, these variables include as special cases the commonly-assumed Gaussian priors for $\bbeta$ in models~\eqref{eq2}--\eqref{eq9}, while extending such distributions in several directions. Hence, this review not only unifies and extends a broad class of models within a single likelihood representation, but also enlarges the class of prior distributions which admit closed-form posteriors that facilitate Bayesian inference.

\subsection{Unified skew-normal prior}\label{sec_3.1}

Routine Bayesian implementations of the models in Sections~\ref{sec_2.1}--\ref{sec_2.4} often assume multivariate Gaussian priors for $\bbeta$, which are natural choices in Bayesian regression and, under the models presented in Sections~\ref{sec_2.1}--\ref{sec_2.4}, are further motivated by the Gaussian form of the underlying latent utilities \citep[e.g.,][]{CHIB199279,Albert_1993, mcculloch1994,nobile1998,Chib1998,mcculloch2000,albert2001,imai2005,kuss2005assessing,holmes_2006,riihimki2012nested,chopin_2017}. Interestingly, these priors are special cases of more general distributions which induce asymmetric shapes in multivariate Gaussians by modifying the density of such variables through a skewness-inducing mechanism driven by the cumulative distribution function of another Gaussian. Key examples include multivariate skew-normals \citep{AZZALINI_dellaValle_1996,azzalini1999}, extended multivariate skew-normals \citep{arnold2000,arnold2002} and closed skew-normals \citep{gonzalez2004,gupta2004}, which have all been subsequently unified by \citet{ARELLANO_VALLE_2006} in a single general class, namely the unified skew-normal  (SUN) distribution. Recalling \citet{ARELLANO_VALLE_2006}, the vector $\bbeta \in \mathbbm{R}^{\bar{p}}$ has $\mbox{SUN}_{\bar{p}, \bar{n}}(\bxi,\bOmega,\bDelta,\bgamma,\bGamma)$ prior  if its density $p(\bbeta)$ is equal to
\begin{equation}
 \phi_{\bar{p}}(\bbeta - \bxi;\bOmega) \frac{\Phi_{\bar{n}}(\bgamma + \bDelta^\intercal \bar{\bOmega}^{-1} \bomega^{-1}(\bbeta-\bxi); \bGamma - \bDelta^\intercal {\bar{\bOmega}^{-1}}\bDelta)}{\Phi_{\bar{n}}(\bgamma; \bGamma)},
    \label{eq11}
\end{equation}
with $\bar{\bOmega}$ denoting the $\bar{p} \times \bar{p}$ correlation matrix associated with the covariance matrix $\bOmega$ which, in turn, can be expressed as $\bOmega=\bomega\bar{\bOmega}\bomega$, where $\bomega=(\bOmega \odot \bI_{\bar{p}})^{1/2}$, and $\odot$ refers to the element-wise Hadamard product. According to \eqref{eq11}, skewness is induced in $\phi_{\bar{p}}(\bbeta - \bxi;\bOmega)$ by multiplying such a density with the cumulative distribution function of a $\mbox{N}_{\bar{n}}({\bf 0},\bGamma - \bDelta^\intercal \bar{\bOmega}{}^{-1} \bDelta)$, evaluated at $\bgamma + \bDelta^\intercal \bar{\bOmega}{}^{-1} \bomega^{-1}(\bbeta-\bxi)$, whereas $\Phi_{\bar{n}}(\bgamma; \bGamma)$ corresponds to the normalizing constant. Notice that when all the entries in the $\bar{p} \times \bar{n}$ skewness matrix $\bDelta$ are $0$, the numerator in \eqref{eq11} reduces to  $\Phi_{\bar{n}}(\bgamma; \bGamma)$, thereby allowing to obtain the classical Gaussian prior density $ \phi_{\bar{p}}(\bbeta - \bxi;\bOmega)$ as a special case of \eqref{eq11}. The quantities $\bar{p}$ and $\bar{n}$ denote instead the dimensions of the density and the cumulative distribution function, respectively. Within the general class of formulations discussed in  Sections~\ref{sec_2.1}--\ref{sec_2.4}, $\bar{p}$ refers to the dimension of $\bbeta$ and, hence, can vary depending on the model considered. While in most cases $\bar{p}$ is equal to the number of predictors $p$, under specific constructions the two dimensions might differ. For instance, in the multinomial probit model  in \eqref{eq7}, $\bar{p}$ coincides with $p\cdot(L-1)$. Conversely, $\bar{n}$ defines the dimension of the multivariate Gaussian cumulative distribution function responsible for the skewness-inducing mechanism in the prior density. For example, setting $\bar{n}=0$ yields the classical Gaussian prior for $\bbeta$, whereas assuming independent skew-normals for each $\beta_j$, $j=1, \ldots, \bar{p}$, would imply $\bar{n}=\bar{p}$.

Recalling \citet{ARELLANO_VALLE_2006}, the above SUN distribution also admits a generative construction which further clarifies the role of the parameters $\bxi$, $\bOmega$, $\bDelta$, $\bgamma$, and $\bGamma$, and provides key intuitions on the conjugacy properties of SUN priors under likelihood \eqref{eq1}. In particular, let $\tilde{\bbeta} \in  \mathbbm{R}^{\bar{p}}$ and $\tilde{\bz} \in  \mathbbm{R}^{\bar{n}}$ denote two vectors jointly distributed as a $\mbox{N}_{\bar{p}+\bar{n}}({\bf 0}, \bOmega^*)$, where $\bOmega^*$ is a $(\bar{p}+\bar{n}) \times (\bar{p}+\bar{n})$ correlation matrix with blocks $\smash{\bOmega^*_{[11]}=\bar{\bOmega}}$, $\smash{\bOmega^*_{[22]}=\bGamma}$ and $\smash{\bOmega^*_{[21]}=\bOmega^{*\intercal}_{[12]}=\bDelta^{\intercal}}$, then $\bar{\bbeta}=(\tilde{\bbeta} \mid \tilde{\bz}+\bgamma>{\bf 0})$ is distributed as a $\mbox{SUN}_{\bar{p}, \bar{n}}({\bf 0},\bar{\bOmega},\bDelta,\bgamma,\bGamma)$, while $\bbeta= \bxi+\bomega \bar{\bbeta} \sim \mbox{SUN}_{\bar{p}, \bar{n}}(\bxi,\bOmega,\bDelta,\bgamma,\bGamma)$ with density as in \eqref{eq11}. Consistent with this generative representation, the parameters $\bxi$ and $\bomega$ control the location and the scale of the prior, whereas $\bar{\bOmega}$, $\bGamma$ and $\bDelta$ regulate the dependence within $\tilde{\bbeta}$, $\tilde{\bz}$ and between these two random vectors, respectively. The term $\bgamma$ denotes instead the truncation threshold in the conditioning mechanism. Besides clarifying the role of the prior parameters, this representation offers intuitions on the SUN conjugacy formalized in Section~\ref{sec_3.2}. In fact, according to such  a construction, SUNs arise as conditional distributions in a generative mechanism that relies on partially-observed Gaussian latent  variables $\tilde{\bz}$. This  has direct connections with the posterior distribution for $\bbeta$ under the models  in Section~\ref{sec_2}, which is also defined, through Bayes rule, via a conditioning operation relying on partially or fully observed Gaussian utilities. 

The above discussion further suggests that different forms of prior information can  be included via \eqref{eq11}. Since multivariate normal distributions are special cases of SUNs, all  non-informative, weakly informative and informative priors relying on Gaussians \citep[e.g.,][]{zellner1986assessing,Gelman2008,chopin_2017} can be employed by letting $\bar{n}=0$, and suitably specifying $\bxi$ and $\bOmega$. The possibility to include skewness further allows the incorporation of additional prior information by letting $\bar{n}>0$ and choosing appropriate values for the parameters $\bDelta$, $\bgamma$ and $\bGamma$, keeping in mind the corresponding role in the generative process that leads to the prior density in \eqref{eq11}. Key examples  of priors of potential interest in this context, which belong to the SUN family, are univariate skew-normals  \citep{azzalini1985} for each coefficient $\beta_j$, $j=1, \ldots, \bar{p}$, or multivariate skew-normals \citep{AZZALINI_dellaValle_1996,azzalini1999}, extended multivariate skew-normals \citep{arnold2000,arnold2002} and closed skew-normals \citep{gonzalez2004,gupta2004} for the entire vector $\bbeta$. Among such options, independent univariate skew-normals  are a convenient choice that easily allows to elicit skewness information for each  $\beta_j$, $j=1, \ldots, \bar{p}$ via a single and  interpretable parameter.

As clarified in Section~\ref{sec_3.2}, the SUN properties are also beneficial for posterior inference. Recalling  \citet{ARELLANO_VALLE_2006,azzalini2010prospective,azzalini_2013}, and \citet{arellano2021} SUNs have a number of  properties in common with multivariate Gaussians. These include closure under marginalization, linear combinations and conditioning, along with the availability of closed-form expressions for the moment generating function, and additive representations via linear combinations of multivariate Gaussians and truncated normals. Due to the SUN conjugacy proved in Theorem~\ref{teo1}, all these properties can facilitate point estimation, uncertainty quantification, model selection and prediction under the SUN posterior associated with the general likelihood in \eqref{eq1} which encompasses the models in Sections~\ref{sec_2.1}--\ref{sec_2.4}. This provides important advancements for a broad class of models, under a similarly wide family of priors beyond  Gaussians. 

\subsection{Unified skew-normal posterior and its properties}\label{sec_3.2}

Theorem~\ref{teo1} unifies and extends recent model-specific derivations by proving SUN conjugacy for any statistical model whose likelihood can be expressed as in~\eqref{eq1}. The proof of Theorem~\ref{teo1} combines original results on SUN conjugacy in probit models \citep[][Corollary 4]{durante_2019} with Lemma~\ref{lem1} below, which shows that SUN priors are also conjugate to Gaussian linear regression; see Section~S2 in the Supplementary Materials for detailed proofs of Lemma~\ref{lem1} and Theorem~\ref{teo1}.

\begin{Lemma}
Let $p(\bar{\by}_1 \mid \bbeta)= \phi_{\bar{n}_1}(\bar{\by}_1-\bar{\bX}_1 \bbeta;\bar{\bSigma}_1)$ and assume that $\bbeta$ has  $\mbox{\normalfont SUN}_{\bar{p}, \bar{n}}(\bxi,\bOmega,\bDelta,\bgamma,\bGamma)$ prior  with density $p(\bbeta)$ as in \eqref{eq11}. Then, $(\bbeta \mid \bar{\by}_{1}) \sim \mbox{\normalfont SUN}_{\bar{p}, \bar{n}}(\bxi_1,\bOmega_1,\bDelta_1,\bgamma_1,\bGamma_1)$ where 
\begin{equation*}
\begin{split}
\bxi_1&=(\bOmega^{-1}+\bar{\bX}^{\intercal}_1 \bar{\bSigma}{}^{-1}_1\bar{\bX}_1)^{-1}(\bOmega^{-1} \bxi+\bar{\bX}^{\intercal}_1 \bar{\bSigma}{}^{-1}_1 \bar{\by}_1), \\
 \bOmega_1&=(\bOmega^{-1}+\bar{\bX}^{\intercal}_1 \bar{\bSigma}{}^{-1}_1\bar{\bX}_1)^{-1}=\bomega_1 \bar{\bOmega}_1\bomega_1, \\
\bDelta_1&= \bar{\bOmega}_1\bomega_1\bomega^{-1} \bar{\bOmega}{}^{-1}\bDelta\bs^{-1}_1,\\
\bgamma_1&=\bs^{-1}_1[\bgamma+\bDelta^\intercal \bar{\bOmega}{}^{-1} \bomega^{-1}(\bxi_1-\bxi)], \\ \bGamma_1&=\bs^{-1}_1[\bGamma+\bDelta^\intercal (\bar{\bOmega}{}^{-1}\bomega^{-1} \bOmega_1\bomega^{-1}\bar{\bOmega}{}^{-1}-\bar{\bOmega}{}^{-1})\bDelta]\bs^{-1}_1,
\end{split}
\end{equation*}
with $\bs_1=([\bGamma{+}\bDelta^\intercal (\bar{\bOmega}^{-1}\bomega^{-1} \bOmega_1\bomega^{-1}\bar{\bOmega}^{-1}{-}\ \bar{\bOmega}^{-1})\bDelta] \odot \bI_{\bar{n}})^{1/2}$.
\label{lem1}
\end{Lemma}

\noindent Note that in Lemma~\ref{lem1} the rescaling operated by $\bs_1$ is required to ensure that the matrix $\bOmega_1^*$ with blocks  $\smash{\bOmega^*_{1[11]}=\bar{\bOmega}_1}$, $\smash{\bOmega^*_{1[22]}=\bGamma_1}$ and $\smash{\bOmega^*_{1[21]}=\bOmega^{*\intercal}_{1[12]}=\bDelta_1^{\intercal}}$ is a correlation matrix, as in the original formulation \citep{ARELLANO_VALLE_2006}. Although this constraint is useful to avoid identifiability issues in frequentist contexts, such  problems are less of a concern in our Bayesian setting since the parameters of the SUN posterior are known functions of the observed data and of the pre-specified prior hyperparameters, and thus, do not need to be estimated. However, maintaining this constraint is still useful to inherit results of the original SUN  and to avoid identifiability issues in prior elicitation.

Leveraging Lemma~\ref{lem1} above, and adapting Corollary 4 in \citet{durante_2019}, it is now possible to state the general SUN conjugacy result in Theorem~\ref{teo1}.

\begin{Theorem}
Consider the likelihood $$p(\by  \mid \bbeta) \propto  \phi_{\bar{n}_1}(\bar{\by}_1-\bar{\bX}_1 \bbeta;\bar{\bSigma}_1)\Phi_{\bar{n}_0}(\bar{\by}_0+\bar{\bX}_0 \bbeta; \bar{\bSigma}_0)$$ defined in \eqref{eq1}, and assume that $\bbeta$ has a $\mbox{\normalfont SUN}_{\bar{p}, \bar{n}}(\bxi,\bOmega,\bDelta,\bgamma,\bGamma)$ prior with density  as in \eqref{eq11}. Then $$(\bbeta \mid {\by}) \sim \mbox{\normalfont SUN}_{\bar{p}, \bar{n}+\bar{n}_0}(\bxi_{\pst},\bOmega_{\pst},\bDelta_{\pst},\bgamma_{\pst},\bGamma_{\pst})$$ with posterior  parameters
\begin{equation*}
\begin{split}
&\bxi_{\pst}=\bxi_1, \quad  \bOmega_{\pst}=\bOmega_1,  \quad  \bDelta_{\pst}= (\bDelta_1, \bar{\bOmega}_{1}\bomega_{1}\bar{\bX}^{\intercal}_0 \bs^{-1}_0),  \\
&  \bgamma_{\pst}=(\bgamma^{\intercal}_1,(\bar{\by}_0 +\bar{\bX}_0\bxi_{1})^{\intercal} \bs^{-1}_0)^{\intercal},
\end{split}
\end{equation*}
and $\bGamma_{\pst}$ denoting a full-rank $(\bar{n}+\bar{n}_0) \times (\bar{n}+\bar{n}_0)$ correlation matrix with blocks 
\begin{equation*}
\begin{split}
& \bGamma_{\pst[11]}=\bGamma_1, \quad \  \bGamma_{\pst[22]}=\bs^{-1}_0(\bar{\bX}_0\bOmega_{1} \bar{\bX}^{\intercal}_0+\bar{\bSigma}_0) \bs^{-1}_0, \\
& \bGamma_{\pst[21]}=\bGamma^{\intercal}_{\pst[12]}=\bs^{-1}_{0}\bar{\bX}_0\bomega_1\bDelta_1,
\end{split}
\end{equation*}
where  $\bs_0=[(\bar{\bX}_0\bOmega_{1} \bar{\bX}^{\intercal}_0+\bar{\bSigma}_0)\odot \bI_{\bar{n}_0}]^{1/2}$, while $\bxi_1$, $\bOmega_1$, $\bDelta_1$, $\bgamma_1$ and $\bGamma_1$ are defined as in Lemma~\ref{lem1}.
\label{teo1}
\end{Theorem}

Theorem~\ref{teo1} encompasses all available conjugacy results for SUN distributions under specific models within the broader family analyzed, while extending such findings to other key formulations. For example, setting $\bar{p}=p$, $\bar{n}_1=0$, $\bar{n}_0=n$, $\bar{\by}_0={\bf 0}$, $\bar{\bX}_0=\mbox{diag}(2\by-{\bf 1}_n)\bX$ and $\bar{\bSigma}_0=\bI_n$ as in \eqref{eq5}, and substituting these quantities within the expressions in Theorem~\ref{teo1}, would yield a  $ \mbox{SUN}_{p, \bar{n}+n}(\bxi_{\pst},\bOmega_{\pst},\bDelta_{\pst},\bgamma_{\pst},\bGamma_{\pst})$ posterior with parameters as in Corollary 4 by \citet{durante_2019}. Theorem 1 in \citet{durante_2019} is instead recovered under the additional constraint $\bar{n}=0$, which implies a Gaussian prior. Note that when $\bar{n}_1=0$ the associated quantities $\bar{\by}_1$, $\bar{\bX}_1$ and $\bar{\bSigma}_1$ are not defined and simply need to be removed from the formulas in Theorem~\ref{teo1}. The same reasoning holds for $\bar{\by}_0$, $\bar{\bX}_0$ and $\bar{\bSigma}_0$ when $\bar{n}_0=0$, and for $\bDelta$, $\bgamma$ and $\bGamma$ if $\bar{n}=0$. For instance, setting $\bar{n}_0=0$ in Theorem~\ref{teo1} leads to Lemma~\ref{lem1}. Similarly, the  SUN conjugacy results for multinomial probit \citep{fasano2020}, dynamic multivariate probit \citep{fasano2021closed}, Gaussian processes \citep{cao2020scalable}, and skewed Gaussian processes under linear models, affine probit and combinations of these two formulations \citep{benavoli2020skew,benavoli2021unified} can be readily obtained from Theorem~\ref{teo1} under the settings in Sections~\ref{sec_2.1}--\ref{sec_2.4} for the quantities defining the likelihood in \eqref{eq1}. Interestingly, also results outside of the regression context, such as those proved by \citet{canale_2016} for multivariate skew-normal likelihoods with Gaussian or skew-normal priors on the shape parameter, can be recasted within Theorem~\ref{teo1}. Besides encompassing already available findings, Theorem~\ref{teo1} provides novel conjugacy results also in previously-unexplored settings, such as in tobit regression and in models relying on skewed utilities.

As discussed in Section~\ref{sec_3.1}, the availability of a SUN posterior in Theorem~\ref{teo1} facilitates Bayesian inference for the whole class of models  in Sections~\ref{sec_2.1}--\ref{sec_2.4}, by leveraging known properties of SUNs \citep[][]{azzalini_2013,arellano2021}. For instance, recalling \citet{ARELLANO_VALLE_2006}, the moment generating function of the posterior is
\begin{eqnarray}
    M(\btt) = e^{\bxi_{\pst}^{\intercal} \btt+0.5\btt^{\intercal}\bOmega_{\pst}\btt}\frac{\Phi_{\bar{n}+\bar{n}_0}(\bgamma_{\pst}+\bDelta^{\intercal}_{\pst}\bomega_{\pst}\btt;\bGamma_{\pst})}{\Phi_{\bar{n}+\bar{n}_0}(\bgamma_{\pst};\bGamma_{\pst})},
    \label{eq12}
\end{eqnarray}
for $\btt \in \mathbbm{R}^{\bar{p}}$, and, therefore, closed-form expressions for relevant moments can be obtained from \eqref{eq12}. In particular, applying the derivations in  \citet{azzalini2010prospective} and \citet{arellano2021} to the SUN posterior in Theorem~\ref{teo1}, yields the following expressions for $\mathbb{E}(\bbeta \mid \by)$ and $\mbox{var}(\bbeta \mid \by)$
\begin{equation}
\begin{split}
   & \mathbb{E}(\bbeta \mid \by)=\bxi_{\pst}+\bomega_{\pst}\bDelta_{\pst}\bpsi, \\ 
   & \mbox{var}(\bbeta \mid \by)=\bOmega_{\pst}+\bomega_{\pst}\bDelta_{\pst}(\bPsi-\bpsi \bpsi^{\intercal})\bDelta^{\intercal}_{\pst}\bomega_{\pst},
    \end{split}
    \label{eq13}
\end{equation}
where $\bpsi$ denotes a vector of dimension $(\bar{n}+\bar{n}_0) \times 1$, with  elements $\psi_i=\phi(\gamma_{\pst,i})\Phi_{\bar{n}+\bar{n}_0-1}(\bgamma_{\pst, -i}-\bGamma_{\pst, -i}\gamma_{\pst,i}; \bGamma_{\pst, -i,-i}-\bGamma_{\pst, -i}\bGamma^{\intercal}_{\pst, -i})/\Phi_{\bar{n}+\bar{n}_0}(\bgamma_{\pst};\bGamma_{\pst})$ for $i=1, \ldots, \bar{n}+\bar{n}_0$, where $\gamma_{\pst,i}$ and $\bgamma_{\pst, -i}$ correspond to the $i$th element of $\bgamma_{\pst}$ and the $(\bar{n}+\bar{n}_0-1) \times 1$ vector obtained by removing entry $i$ in $\bgamma_{\pst}$, respectively, whereas $\bGamma_{\pst, -i}$ and $\bGamma_{\pst, -i,-i}$ are the $i$th column of $\bGamma_{\pst}$ without entry $i$ and the sub-matrix obtained by removing the $i$th row and column from $\bGamma_{\pst}$, respectively. Analogously, $\bPsi$ is a $(\bar{n}+\bar{n}_0) \times (\bar{n}+\bar{n}_0)$ symmetric matrix involving the second-order derivatives of the cumulative distribution function term in \eqref{eq12}; refer to \citet{arellano2021} for the exact expression of $\bPsi$ and of higher-order moments of the SUN. These quantities can be also estimated via Monte Carlo since
\begin{eqnarray}
(\bbeta \mid \by) \stackrel{\mbox{\small d}}{=} \bxi_{\pst}+\bomega_{\pst}(\bU_0+\bDelta_{\pst}\bGamma^{-1}_{\pst}\bU_1),
    \label{eq14.add}
\end{eqnarray}
with $\stackrel{\mbox{\small d}}{=}$ meaning equality in distribution, and
\begin{equation*}
\begin{split}
&\bU_0 \sim \mbox{N}_{\bar{p}}({\bf 0}, \bar{\bOmega}_{\pst}-\bDelta_{\pst}\bGamma^{-1}_{\pst}\bDelta^{\intercal}_{\pst}), \\
&\bU_1 \sim \mbox{TN}_{\bar{n}+\bar{n}_0}(-\bgamma_{\pst}; {\bf0}, \bGamma_{\pst}),
\end{split}
\end{equation*}
where  $\mbox{TN}_{\bar{n}+\bar{n}_0}(-\bgamma_{\pst}; {\bf0}, \bGamma_{\pst})$ denotes an $(\bar{n}+\bar{n}_0)$--variate Gaussian having mean ${\bf 0}$, covariance matrix $ \bGamma_{\pst}$ and truncation below $-\bgamma_{\pst}$. This additive construction has been first derived in \citet{ARELLANO_VALLE_2006} and allows to generate independent and identically distributed values from the exact posterior via linear combinations of samples from $\bar{p}$-variate Gaussians and $(\bar{n}+\bar{n}_0)$-variate truncated normals, thus overcoming convergence and mixing issues of MCMC methods; see Section~\ref{sec_4}.

Uncertainty quantification and calculation of credible intervals is instead facilitated by the availability of a closed-form expression for the SUN cumulative distribution function. Adapting \citet{azzalini2010prospective} and \citet{arellano2021}, this is
\begin{equation}
\mbox{pr}(\bbeta \leq \bb \mid \by)= \frac{\Phi_{\bar{p}+(\bar{n}+\bar{n}_0)}([(\bb-\bxi_{\pst})^{\intercal}\bomega^{-1}_{\pst},\bgamma^{\intercal}_{\pst}]^{\intercal};\tilde{\bOmega}_{\pst})}{\Phi_{\bar{n}+\bar{n}_0}(\bgamma_{\pst};\bGamma_{\pst})}, 
    \label{eq14}
\end{equation}
for  $\bb \in  \mathbbm{R}^{\bar{p}}$, where $\tilde{\bOmega}_{\pst}$  is a matrix with blocks $\smash{\tilde{\bOmega}_{\pst[11]}=\bar{\bOmega}_{\pst}}$, $\tilde{\bOmega}_{\pst[22]}={\bGamma}_{\pst}$ and \smash{$\tilde{\bOmega}_{\pst[21]}=\tilde{\bOmega}^{\intercal}_{\pst[12]}=-{\bDelta}^{\intercal}_{\pst}$}.

Extending the results of \citet{durante_2019}, \citet{fasano2020} and  \citet{benavoli2021unified} to the more general setting under consideration, it is also possible to obtain the marginal likelihood as follows
\begin{equation}
p(\by)=c \cdot \phi_{\bar{n}_1}(\bar{\by}_1-\bar{\bX}_1 \bxi; \bar{\bSigma}_{1}+\bar{\bX}_1\bOmega \bar{\bX}^{\intercal}_1)\frac{\Phi_{\bar{n}+\bar{n}_0}(\bgamma_{\pst};\bGamma_{\pst})}{\Phi_{\bar{n}}(\bgamma;\bGamma)},
    \label{eq15}
\end{equation}
where $c=1$ under all the routinely-implemented models in Section~\ref{sec_2} which rely on Gaussian utilities, namely \eqref{eq2}--\eqref{eq9}, whereas for those formulations based on skewed utilities, e.g., (S.1)--(S.3), the constant $c$ is a known value. Albeit the primary interest is  inference on  $\bbeta$, the marginal likelihood in \eqref{eq15} allows to obtain empirical Bayes estimates also for the other quantities in likelihood \eqref{eq1}, such as the parameters of the covariance matrices $\bar{\bSigma}_{1}$ and $\bar{\bSigma}_{0}$, via numerical maximization; see Section~\ref{sec_6} for further discussion on estimation of $\bar{\bSigma}_{1}$ and $\bar{\bSigma}_{0}$. In addition, Equation~\eqref{eq15} facilitates direct calculation of Bayes factors for model selection and evaluation of predictive probabilities. This second objective can be readily accomplished noting that the predictive probability $p(\by_{\mbox{\normalfont \tiny new}} \mid \by)$ for a new vector of observations $\by_{\mbox{\normalfont \tiny new}}$ from model \eqref{eq1} is equal to the ratio $p(\by_{\mbox{\normalfont \tiny new}}, \by)/p(\by)$ of the two associated marginal likelihoods. Therefore, focusing for simplicity on the case $c=1$, that covers the most widely--used models in Section~\ref{sec_2},  direct application of \eqref{eq15} leads to
\begin{equation}
\begin{split}
p(\by_{\mbox{\normalfont \tiny new}} \mid \by)=& \ \frac{\phi_{\bar{n}_1+\bar{n}_{1\mbox{\normalfont \tiny new}}}(\bar{\by}_{1\mbox{\normalfont \tiny pred}}-\bar{\bX}_{1\mbox{\normalfont \tiny pred}} \bxi; \bar{\bSigma}_{1\mbox{\normalfont \tiny pred}}+\bar{\bX}_{1\mbox{\normalfont \tiny pred}}\bOmega \bar{\bX}^{\intercal}_{1\mbox{\normalfont \tiny pred}})}{\phi_{\bar{n}_1}(\bar{\by}_1-\bar{\bX}_1 \bxi; \bar{\bSigma}_{1}+\bar{\bX}_1\bOmega \bar{\bX}^{\intercal}_1)} \\
& \ \ \cdot \frac{\Phi_{\bar{n}+\bar{n}_0+\bar{n}_{0\mbox{\normalfont \tiny new}}}(\bgamma_{\mbox{\normalfont \tiny pred}};\bGamma_{\mbox{\normalfont \tiny pred}})}{\Phi_{\bar{n}+\bar{n}_0}(\bgamma_{\pst};\bGamma_{\pst})},
\end{split}
 \label{eq16}
\end{equation}
where $\bar{n}_{1\mbox{\normalfont \tiny new}}$ and $\bar{n}_{0\mbox{\normalfont \tiny new}}$ are the dimensions of the two vectors $\bar{\by}_{1\mbox{\normalfont \tiny new}}$ and $\bar{\by}_{0\mbox{\normalfont \tiny new}}$ associated with $\by_{\mbox{\normalfont \tiny new}}$. Similarly, $\bar{\by}_{1\mbox{\normalfont \tiny pred}}=(\bar{\by}^{\intercal}_1,\bar{\by}^{\intercal}_{1\mbox{\normalfont \tiny new}})^{\intercal}$, $\bar{\bX}_{1\mbox{\normalfont \tiny pred}}=(\bar{\bX}^{\intercal}_1,\bar{\bX}^{\intercal}_{1\mbox{\normalfont \tiny new}})^{\intercal} $, whereas $\bar{\bSigma}_{1\mbox{\normalfont \tiny pred}}$ denotes the block-diagonal matrix having $\bar{\bSigma}_{1\mbox{\normalfont \tiny pred}[11]}=\bar{\bSigma}_1$ and $\bar{\bSigma}_{1\mbox{\normalfont \tiny pred}[22]}=\bar{\bSigma}_{1\mbox{\normalfont \tiny new}}$. The two quantities $\bgamma_{\mbox{\normalfont \tiny pred}}$ and $\bGamma_{\mbox{\normalfont \tiny pred}}$, and, implicitly, $\bxi_{\mbox{\normalfont \tiny pred}}$, $\bOmega_{\mbox{\normalfont \tiny pred}}$ and $\bDelta_{\mbox{\normalfont \tiny pred}}$, are constructed analogously to the posterior parameters in Theorem~\ref{teo1}, after replacing the original data with the enriched ones $(\bar{\by}_{1\mbox{\normalfont \tiny pred}},\bar{\bX}_{1\mbox{\normalfont \tiny pred}},\bar{\bSigma}_{1\mbox{\normalfont \tiny pred}})$ and $(\bar{\by}_{0\mbox{\normalfont \tiny pred}},\bar{\bX}_{0\mbox{\normalfont \tiny pred}},\bar{\bSigma}_{0\mbox{\normalfont \tiny pred}})$, where $\bar{\by}_{0\mbox{\normalfont \tiny pred}}=(\bar{\by}^{\intercal}_0,\bar{\by}^{\intercal}_{0\mbox{\normalfont \tiny new}})^{\intercal}$, $\bar{\bX}_{0\mbox{\normalfont \tiny pred}}=(\bar{\bX}^{\intercal}_0,\bar{\bX}^{\intercal}_{0\mbox{\normalfont \tiny new}})^{\intercal} $ and $\bar{\bSigma}_{0\mbox{\normalfont \tiny pred}}$ is a block--diagonal matrix with $\bar{\bSigma}_{0\mbox{\normalfont \tiny pred}[11]}=\bar{\bSigma}_0$ and $\bar{\bSigma}_{0\mbox{\normalfont \tiny pred}[22]}=\bar{\bSigma}_{0\mbox{\normalfont \tiny new}}$.

Before concluding the overview of the SUN properties that facilitate posterior inference, it shall be emphasized that  SUNs are closed under marginalization, linear combinations and conditioning \citep{ARELLANO_VALLE_2006,arellano2021}. This means, for instance, that the posterior distribution of any sub-vector $\bbeta_{[{\bf j}]}$, ${\bf j} \subset \{1, \ldots , \bar{p}\}$ is $\mbox{SUN}_{|{\bf j}|, \bar{n}+\bar{n}_0}(\bxi_{\pst[{\bf j}]},\bOmega_{\pst[{\bf j}{\bf j}]},\bDelta_{\pst[{\bf j}]},\bgamma_{\pst},\bGamma_{\pst})$, where $\bDelta_{\pst[{\bf j}]}$ corresponds to the matrix $\bDelta_{\pst}$ after deleting all the rows with indexes  not in ${\bf j}$. Therefore, setting ${\bf j}=\{j\}$ shows that the posterior of each $\beta_j$, $j=1, \ldots, \bar{p}$ is still a SUN. Similarly, the posterior distribution for the linear combination $(\ba+\bA^{\intercal} \bbeta) \in \mathbbm{R}^d$ is $\mbox{SUN}_{d, \bar{n}+\bar{n}_0}(\ba{+}\bA^{\intercal}\bxi_{\pst},\bA^{\intercal}\bOmega_{\pst}\bA{,}[{(}\bA^{\intercal}\bOmega_{\pst}\bA{)} \odot \bI_{d}]^{-1/2} \bA^{\intercal}\bomega_{\pst}\bDelta_{\pst},\bgamma_{\pst},\bGamma_{\pst})$. In particular, this implies that the posterior distribution of any linear predictor is still SUN.

The results presented in this section also clarify that, unlike for Bayesian linear regression with Gaussian priors, it is not immediate to disentangle the role of the prior parameters from the one of the data in the functionals and  shape of the SUN posterior  treated in this article. In fact, as clarified in~\eqref{eq14.add}, each of these quantities covers multiple roles in controlling location, scale and skewness; see \citet{durante_2019} for an attempt to separate the effect of the different terms in probit regression with Gaussian priors.

\section{Computational methods}\label{sec_4}

The results presented in Section~\ref{sec_3.2} suggest that posterior inference under the models illustrated in Section~\ref{sec_2} can be performed via closed-form solutions. This is true for any, even huge, $\bar{p}$ as long as  $\bar{n}+\bar{n}_0$ is small-to-moderate, but not when $\bar{n}+\bar{n}_0$ exceeds few hundreds \citep[][]{durante_2019,fasano2020}. In fact, Equations \eqref{eq12}--\eqref{eq16} require evaluation of cumulative distribution functions of $(\bar{n}+\bar{n}_0)$-variate Gaussians or sampling from $(\bar{n}+\bar{n}_0)$-variate truncated normals, which is known to be computationally challenging in high dimensions \citep{Genz1992NumericalCO,GenzBretz2009,Botev_2016,Genton2018,cao2019hierarchical,Cao2020}. This motivates still active research on developing sampling-based methods and accurate deterministic approximations for tractable Bayesian inference under the models  in Section~\ref{sec_2}.  Sections~\ref{sec_4.1}--\ref{sec_4.3} review, unify, extend and compare both past and more recent developments along these lines.

\subsection{Analytical methods}\label{sec_4.1}

As discussed above, the evaluation of high-dimensional Gaussian integrals with linear constraints, such as those found in Equations \eqref{eq12}--\eqref{eq16}, is a longstanding problem \citep[e.g.,][]{Genz1992NumericalCO,genz2002comparison,miwa2003evaluation,gassmann2003multivariate,genz2004numerical,craig2008new,ridgway2016,Botev_2016,Genton2018,cao2019hierarchical,gessner2020,Cao2020}.  

A popular class of strategies for evaluating these Gaussian integrals encompasses several extensions of the original  separation of variables estimator initially proposed by \citet{Genz1992NumericalCO}. This solution recasts the problem as a sequence of tractable one-dimensional integrals, which are evaluated numerically via a randomized quasi-Monte Carlo sampling.  As suggested in, e.g., \citet{GenzBretz2009}, the variance of the resulting estimator can be further reduced by  means of variable reordering. More recently, \citet{Botev_2016} proposed a new solution  relying on an  optimal exponential tilting of the \citet{Genz1992NumericalCO} construction, which is found by solving efficiently a minimax saddle-point problem, and then used as an effective importance sampling proposal. While still providing an unbiased estimate, this technique achieves  practical reduction of the estimator variance by orders of magnitude.  Moreover, this procedure remains effective in settings where the  \citet{Genz1992NumericalCO}  method cannot provide reliable estimates. Such a solution, available in the \texttt{R} library \texttt{TruncatedNormal}, remains generally tractable in a few hundreds of dimensions, but it progressively slows down beyond this regime. To achieve scalability in higher dimensions, recent solutions leverage low-rank hierarchical  block structures of the covariance matrix within the high-dimensional Gaussian integral to decompose the problem into a sequence of smaller-dimensional ones which facilitate reduction of computational cost while preserving accuracy \citep{Genton2018,cao2019hierarchical,Cao2020}. Among these alternatives, the one proposed in \citet{Cao2020} provides a state-of-the-art extension of the original separation of variables estimator which incorporates both an effective tile-low-rank representation of the covariance matrix and an iterative block-reordering scheme to obtain notable improvements in runtimes and scalability. For instance, such a solution has been recently adapted to the problem of evaluating predictive probabilities in high-dimensional probit Gaussian processes with $\bar{n}_0$ and $\bar{p}$ in tens of thousands \citep{cao2020scalable}, obtaining remarkable improvements over state-of-the-art methods.

There are also alternative solutions beyond  the classical separation of variables technique. For example, \citet{ridgway2016} developed a sequential Monte Carlo sampler to compute Gaussian orthant probabilities, adding a dimension at each step, combined with carefully-designed MCMC moves. More recently, \citet{gessner2020} constructed an efficient estimator of Gaussian integrals with linear domain constraints, that decomposes  the problem into a sequence of easier-to-solve conditional probabilities, based on nested domains. 
Each internal step uses an analytic version of elliptical slice sampling, exploiting the availability of closed-form solutions for the intersections between the ellipses and linear constraints.
The authors reported evidence of effectiveness of such method even for thousands-dimensional integrals. Further strategies can be found in, e.g.,  \citet{genz2002comparison,miwa2003evaluation,gassmann2003multivariate,genz2004numerical,craig2008new,trinh2015bivariate}.

Interestingly, some of the aforementioned strategies also provide, as a byproduct, effective solutions for sampling from multivariate truncated normals, which can be useful to generate values from the SUN posterior via the additive representation in~\eqref{eq14.add}. These methods can be found, for example, in  \citet{Botev_2016} and in \citet{gessner2020}. Motivated by inference on a phylogenetic multivariate probit model, \cite{Zhang2019}  recently employed an alternative scheme for sampling from truncated normals with dimension above ten thousands, via a bouncy particle sampler. See also  \citet{PakmanPaninski} for an Hamiltonian Monte Carlo scheme, incorporating the truncations via hard walls and exploiting the possibility to integrate exactly the  Hamiltonian equations.

All the above solutions provide effective methods for evaluating Gaussian cumulative distribution functions and, possibly, sampling from multivariate truncated normals. However, such procedures are still subject to a tradeoff between accuracy and computational tractability which is often specific to the model analyzed and to the size of the data, thereby motivating still ongoing research. Due to this, it is difficult to identify a generally-applicable gold-standard among the aforementioned techniques, although, in practice, the method by \citet{Botev_2016} has often notable performance when applied to Equations \eqref{eq12}--\eqref{eq16} in small-to-moderate size settings with  $\bar{n}+\bar{n}_0$ in the order of few hundreds. Higher-dimensional problems may require more scalable solutions \citep[e.g.,][]{gessner2020,Cao2020,Zhang2019}, even if more extensive empirical analyses are required to assess these methods.

\subsection{Sampling-based methods}\label{sec_4.2}

Whenever the interest is on more complex posterior functionals beyond those derived in Section~\ref{sec_3.2}, an effective solution is to consider Monte Carlo estimates based on samples from $p(\bbeta \mid \by)$. While generally-applicable MCMC strategies such as  state-of-the-art implementations of Hamiltonian Monte Carlo \citep[e.g.,][]{hoffman2014} and Metropolis--Hastings \citep[e.g.,][]{roberts2001} can be considered, a  widely-implemented class of algorithms within the context of the models presented in Section~\ref{sec_2} are data augmentation Gibbs samplers \citep[see e.g.,][]{CHIB199279,Albert_1993,mcculloch1994,Chib1998,albert2001,imai2005,holmes_2006}. This is because  the formulations  in Section~\ref{sec_2} rely on Gaussian latent utilities which are assigned a regression model with coefficients $\bbeta$. Therefore, treating these utilities as augmented data  restores Gaussian-Gaussian conjugacy between the prior for $\bbeta$ and the likelihood of the augmented utilities, which can be in turn sampled from independent truncated normal full-conditionals, given $\bbeta$ and the censoring information provided by the observed $\by$. This yields tractable Gibbs samplers that iterate among these two steps, thus producing samples from the posterior of $\bbeta$.

Although the above techniques have been proposed only for a subset of the models presented in Section~\ref{sec_2}, and in separate contributions mainly focusing on Gaussian priors \citep[e.g.,][]{CHIB199279,Albert_1993,mcculloch1994,holmes_2006}, the comprehensive framework in Equation \eqref{eq1}, and the general conjugacy results reported in Section~\ref{sec_3} allow to unify these  MCMC strategies within a broad construction which can be applied to any model in Section~\ref{sec_2}, even beyond those currently studied, and holds not only for Gaussian prior distributions, but also for the general SUN ones. Letting 
\begin{equation*}
\begin{split}
&\bX_{\pst}=\bDelta^{\intercal}_{\pst}\bar{\bOmega}^{-1}_{\pst}{\bomega}^{-1}_{\pst}, \qquad \boeta_{\pst}=\bgamma_{\pst}-\bX_{\pst}\bxi_{\pst}, \\
&\bSigma_{\pst}=\bGamma_{\pst}-\bDelta^{\intercal}_{\pst}\bar{\bOmega}{}^{-1}_{\pst}\bDelta_{\pst},
\end{split}
\end{equation*}
this general Gibbs sampler can be obtained by noticing that, due to \eqref{eq11}, the density kernel of the SUN posterior in Theorem~\ref{teo1} coincides with $\phi_{\bar{p}}(\bbeta - \bxi_{\pst};\bOmega_{\pst}) \Phi_{\bar{n}+\bar{n}_0}(\boeta_{\pst} + \bX_{\pst}\bbeta; \bSigma_{\pst})$, where the cumulative distribution function term   can be also written as $\int \phi_{\bar{n}+\bar{n}_0}(\bar{\bz}-(\boeta_{\pst} + \bX_{\pst} \bbeta); \bSigma_{\pst}) \mathbbm{1}(\bar{\bz} > {\bf 0}) \mbox{d}\bar{\bz}$. Hence, extending the augmented-data representation by \citet{Albert_1993} --- see also \citet{fasano2020} --- and  leveraging standard properties of multivariate Gaussian and truncated normals, this alternative formulation implies a generally-applicable data augmentation Gibbs sampler relying on the full-conditional distributions
\begin{equation}
\begin{split}
  & (\bbeta \mid \by, \bar{\bz})  \sim \mbox{N}_{\bar{p}}(\bV_{\pst}[\bX^{\intercal}_{\pst} \bSigma^{-1}_{\pst}(\bar{\bz}- \boeta_{\pst})+{\bOmega}^{-1}_{\pst} \bxi_{\pst}], \bV_{\pst}),\\
    & ( \bar{\bz} \mid \by, \bbeta)  \sim \mbox{TN}_{\bar{n}+\bar{n}_0}({\bf 0}; \boeta_{\pst} + \bX_{\pst} \bbeta,\bSigma_{\pst}),
    \label{eq17}
    \end{split}
\end{equation}
where $\bV_{\pst}=({\bOmega}^{-1}_{\pst}+\bX^{\intercal}_{\pst} \bSigma^{-1}_{\pst}\bX^{\intercal}_{\pst})^{-1}$. Hence, available Gibbs samplers for specific models within \eqref{eq1} and yet unexplored extensions to the whole class under general SUN priors, can be  readily obtained as special cases of \eqref{eq17} under suitable specification of the posterior parameters defining the above full-conditionals. It shall also be emphasized that the sampling from the  $(\bar{n}+\bar{n}_0)$-dimensional truncated normal distribution in \eqref{eq17} is usually simplified by the conditional independence properties among the latent utilities underlying most of the regression models presented in Sections~\ref{sec_2.1}--\ref{sec_2.4}. This means that $\bSigma_{\pst}$ is either diagonal or block-diagonal, often with small-dimensional blocks, and, therefore, sampling from $(\bar{\bz} \mid \by, \bbeta)$  simply requires to draw values from univariate or low-dimensional truncated normals. Nonetheless, as discussed in \citet{Johndrow_2018} the dependence structure between $\bbeta$ and $\bar{\bz}$ can still yield to poor mixing; see also \citet{qin2019} for detailed convergence analysis. 

An effective option to obviate the above mixing issues is to sample i.i.d.\ values from the joint posterior $p(\bbeta, \bar{\bz} \mid \by)$, instead of autocorrelated ones as in \eqref{eq17}. Extending the derivations by \citet{holmes_2006} to the whole class of models in \eqref{eq1}, under SUN priors \eqref{eq11}, this task can be accomplished by noting that $p(\bbeta, \bar{\bz} \mid \by)=p(\bbeta \mid \by, \bar{\bz})p(\bar{\bz}\mid \by)$, where $p(\bbeta \mid \by, \bar{\bz})$ coincides with the density of the Gaussian in \eqref{eq17}, whereas $p(\bar{\bz}\mid \by)$ is obtained by marginalizing out from the truncated normal in \eqref{eq17} the $\bbeta$ vector with density $\phi_{\bar{p}}(\bbeta - \bxi_{\pst};\bOmega_{\pst})$. Leveraging standard properties of Gaussian and truncated normal random variables, and recalling \citet{holmes_2006}, this implies that
\vspace{-3pt}
\begin{eqnarray}
( \bar{\bz} \mid \by)  \sim \mbox{TN}_{\bar{n}+\bar{n}_0}({\bf 0}; \bgamma_{\pst},\bGamma_{\pst}).
    \label{eq18}
\end{eqnarray}
Replacing the full-conditional multivariate truncated normal in \eqref{eq17} with the one in  \eqref{eq18}, yields to a scheme for sampling i.i.d.\ values from  $p(\bbeta, \bar{\bz} \mid \by)$ and, as a direct consequence, from the posterior $p(\bbeta \mid \by)$ of interest. To do this, it is sufficient to draw $\bar{\bz}$ from \eqref{eq18} and then generate a value for $\bbeta$ by sampling from the Gaussian in  \eqref{eq17} with mean evaluated at the sampled value of $\bar{\bz}$. This routine is closely related  to the i.i.d.\ sampler based on the additive representation of the SUN in Equation \eqref{eq14.add} that relies on a linear combination among samples from $\bar{p}$-variate Gaussians and $(\bar{n}+\bar{n}_0)$-variate truncated normals \citep{durante_2019, fasano2020, fasano2021closed}. 

Although the above strategies effectively address the mixing and convergence issues of the Gibbs sampler in  \eqref{eq17}, the multivariate truncated normal in \eqref{eq18} is often more challenging from a computational perspective relative to the one in  \eqref{eq17}. In fact, marginalizing out $\bbeta$ in $\mbox{TN}_{\bar{n}+\bar{n}_0}({\bf 0}; \boeta_{\pst} + \bX_{\pst} \bbeta,\bSigma_{\pst})$ induces dependence among the latent utilities in $\bar{\bz}$. This means that, unlike for $\bSigma_{\pst}$, the covariance matrix $\bGamma_{\pst}$ of the truncated normal in \eqref{eq18} has no more a diagonal or block-diagonal structure and, hence, $p(\bar{\bz} \mid \by)$ does not factorize as the product of univariate or low-dimensional truncated normals as for $p(\bar{\bz} \mid \by, \bbeta)$  in  \eqref{eq17}, making the sampling from \eqref{eq18}  more challenging when $(\bar{n}+\bar{n}_0)$ is large. In probit regression, \citet{holmes_2006} address such issue by leveraging the closure under conditioning properties discussed, e.g., in \citet{Horrace2005SomeRO} to sample iteratively from the univariate truncated normal full-conditionals $p(\bar{z}_i \mid \bar{\bz}_{-i} , \by)$, for $i=1, \ldots, \bar{n}+\bar{n}_0$. However, this strategy implies a Gibbs-sampling routine which may be still subject to mixing issues. Alternatively, it is possible to sample directly from $p(\bar{\bz} \mid \by)$ in \eqref{eq18} leveraging the state-of-the-art schemes presented in Section~\ref{sec_4.1} \citep[e.g.,][]{Botev_2016,gessner2020}. However, there is still the lack of a generally-applicable gold-standard for any size of $\bar{p}$ and $\bar{n}+\bar{n}_0$. 

\subsection{Deterministic approximation-based methods}\label{sec_4.3}

Even resorting to state-of-the-art solutions, sampling from the posterior distribution is often prohibitive for  high-dimensional datasets and  large sample sizes  \citep[e.g.,][]{chopin_2017}. In these scenarios, an effective solution is to consider deterministic approximations of the exact posterior. Sections~\ref{sec_4.3.1}--\ref{sec_4.3.2} provide  a unified treatment of classical and more recent VB  \citep{blei_2017} and EP \citep{Minka2001} approximations which are widely-implemented solutions in the context of the models considered in this article; see \citet{chopin_2017} for a review of alternative methods, such as Laplace approximation and INLA \citep{inla_paper}. A detailed derivation and discussion of the computational costs can be found in Section~S3 of the Supplementary Materials.

\subsubsection{Variational Bayes (VB)}  \label{sec_4.3.1}

VB solves a constrained optimization problem that aims at finding the approximating density which is the closest, in  Kullback--Leiber (KL) divergence  \citep{Kullback-Leibler}, to the exact posterior, among all the densities within a pre-specified tractable family  facilitating Bayesian inference. Recalling  \citet{blei_2017}, within the context of models admitting conditionally conjugate constructions with global parameters $\bbeta$ and local augmented data $\bar{\bz}$ --- such as for the formulations in Section~\ref{sec_2} --- the solution of the optimization problem often benefits from taking $p(\bbeta, \bar{\bz} \mid \by)$ as the target density to be approximated, which in turn would yield to an approximation for  $p(\bbeta \mid \by)$ after marginalizing out $\bar{\bz}$ \citep{girolami2006,consonni_2007, fasano2019,fasano2020}. As for the choice of the approximating family $\mathcal{Q}$,  classical solutions \citep[e.g.,][]{girolami2006,consonni_2007} rely on the mean-field assumption \citep[e.g.,][]{blei_2017} which can be generally expressed as $\mathcal{Q}_{\textsc{mf}}=\{ q(\bbeta, \bar{\bz}): q(\bbeta, \bar{\bz})=q(\bbeta) \prod_{c=1}^{C} q(\bar{\bz}_c)\}$, where $\bar{\bz}_1, \ldots, \bar{\bz}_C$ denote distinct sub-vectors of $\bar{\bz}$, such that $\bar{\bz}=(\bar{\bz}^{\intercal}_1, \ldots, \bar{\bz}^{\intercal}_C)^{\intercal}$. Note that the choice of how to factorize $q(\bar{\bz})$ in $C$ independent blocks is often guided by the dependence structures in $\bar{\bz}$. For instance, in models relying on conditionally independent latent utilities, such as those in Section~\ref{sec_2}, it is common to factorize  $q(\bar{\bz})$ consistent with these conditionally independent  sub-vectors. In fact, as illustrated in the context of probit \citep[e.g.,][]{consonni_2007} and multinomial probit  \citep[e.g.,][]{girolami2006}, even without assuming a specific factorization for  $q(\bar{\bz})$, i.e., $C=1$, the optimum $q_{\textsc{mf}}^*(\bar{\bz})$ within the class $\mathcal{Q}_{\textsc{mf}}$ would still factorize as $\prod_{c=1}^{C} q_{\textsc{mf}}^*(\bar{\bz}_c)$, where $\bar{\bz}_1, \ldots, \bar{\bz}_C$ correspond to the subsets of conditionally independent utilities, as implied by the chosen model and prior.

Summarizing the above discussion, the mean-field variational Bayes (MF-VB) solution can be formalized as
\begin{equation}
\begin{split}
q_{\textsc{mf}}^*(\bbeta, \bar{\bz})&=\argmin_{q(\bbeta, \bar{\bz}) \in \mathcal{Q}_{\textsc{mf}}} \mbox{KL}[q(\bbeta, \bar{\bz}) || p(\bbeta, \bar{\bz} \mid \by)]\\
&=\argmax_{q(\bbeta, \bar{\bz}) \in \mathcal{Q}_{\textsc{mf}}} \mbox{ELBO}[q(\bbeta, \bar{\bz})],
    \end{split}
    \label{eq19}
\end{equation}
since $ \mbox{ELBO}[q(\bbeta, \bar{\bz})]=-\mbox{KL}[q(\bbeta, \bar{\bz}) || p(\bbeta, \bar{\bz} \mid \by)]+\log p(\by)$. Recalling \citet{blei_2017},  \eqref{eq19} can be solved using a tractable coordinate ascent variational inference (CAVI) scheme that iteratively updates the solution of the approximating densities for $\bbeta$ and $\bar{\bz}$ via $q^{(t)}_{\textsc{mf}}(\bbeta) \propto \exp \{ \smash{\mathbbm{E}_{q(\bar{\bz})}}[\log p(\bbeta \mid \by, \bar{\bz})] \}$ and \smash{$q^{(t)}_{\textsc{mf}}(\bar{\bz}_c) \propto \exp \{ \mathbbm{E}_{q(\bbeta,\bar{\bz}_{-c})}[\log p(\bar{\bz}_c {\mid} \by, \bbeta,\bar{\bz}_{-c})] \}$}, for $c=1, \ldots, C$, where $\bar{\bz}_{-c}$ coincides with $\bar{\bz}$ without the sub-vector  $\bar{\bz}_{c}$, while the expectation is taken with respect to the most recent update of the variational density over the other conditioning variables. Replacing the full-conditional distributions in these expressions with those in  \eqref{eq17}, and leveraging the closure under conditioning  of multivariate truncated normals \citep{Horrace2005SomeRO}, yields  a general MF-VB  that extends \citet{girolami2006} and \citet{consonni_2007} to the whole class of models and priors presented in Sections~\ref{sec_2}--\ref{sec_3}, and can be obtained via closed-form CAVI updates. More specifically, let $ \bar{\boeta}_{\pst}=\mathbbm{E}_{q(\bar{\bz})}(\bar{\bz})- \boeta_{\pst}$, where $\mathbbm{E}_{q(\bar{\bz})}(\bar{\bz})=[\mathbbm{E}_{q(\bar{\bz}_1)}(\bar{\bz}^{\intercal}_1), \ldots, \mathbbm{E}_{q(\bar{\bz}_C)}(\bar{\bz}^{\intercal}_C)]^{\intercal}$, and define $\bSigma_{\pst(c)}=\bSigma_{\pst[c,c]}- \bSigma_{\pst[c,-c]}(\bSigma_{\pst[-c,-c]})^{-1}\bSigma_{\pst[-c,c]}$, with $ \bSigma_{\pst[c,c]}$,  $\bSigma_{\pst[c,-c]}$, $\bSigma_{\pst[-c,c]}$, and $\bSigma_{\pst[c,c]}$, corresponding to the four blocks of $\bSigma_{\pst}$ when partitioned to highlight the sub-vector $\bar{\bz}_c$ against all the others in $\bar{\bz}_{-c}$. Then, the CAVI updates for  MF-VB are given by
\begin{equation}
\begin{split}
& q^{(t)}_{\textsc{mf}}(\bbeta) =  \phi_{\bar{p}}(\bbeta-\bV_{\pst}(\bX^{\intercal}_{\pst} \bSigma^{-1}_{\pst} \bar{\boeta}_{\pst}+{\bOmega}^{-1}_{\pst} \bxi_{\pst}); \bV_{\pst}), \\
& q^{(t)}_{\textsc{mf}}(\bar{\bz}_c) \propto \phi_{{n}_c}(\bar{\bz}_c- \mathbbm{E}_{q(\bbeta, \bar{\bz}_{-c})}(\bmu_c); \bSigma_{\pst(c)}) \mathbbm{1}(\bar{\bz}_c > {\bf 0}),\\
& \qquad  \qquad  \qquad \qquad  \qquad  \qquad  \qquad \quad  \ \ \mbox{for} \ c=1, \ldots, C,
    \end{split}
    \label{eq20}
\end{equation}
where $n_c$ is  the dimension of $\bar{\bz}_c$, while  $ \mathbbm{E}_{q(\bbeta, \bar{\bz}_{-c})}(\bmu_c)=\smash{\boeta_{\pst[c]}} + \bX_{\pst[c]}  \mathbbm{E}_{q(\bbeta)}(\bbeta)+\bSigma_{\pst[c,-c]}(\bSigma_{\pst[-c,-c]})^{-1}[ \mathbbm{E}_{q(\bar{\bz}_{-c})}(\bar{\bz}_{-c})-\boeta_{\pst[-c]} - \bX_{\pst[-c]}  \mathbbm{E}_{q(\bbeta)}(\bbeta)]$.  The quantities $\bX_{\pst[c]}$, $\bX_{\pst[-c]}$, $\boeta_{\pst[c]}$ and $\boeta_{\pst[-c]}$ within \eqref{eq20} denote the rows of  $\bX_{\pst}$ and  $\boeta_{\pst}$ corresponding to $\bar{\bz}_c$ and $\bar{\bz}_{-c}$, respectively. Hence, according to  \eqref{eq20}, MF-VB for the whole class of models and priors in Sections~\ref{sec_2}--\ref{sec_3} can be implemented via a simple CAVI routine providing Gaussian and truncated normal approximating densities for $\bbeta$ and $\bar{\bz}_1, \ldots, \bar{\bz}_C$, respectively, which only require updating of the corresponding means with respect to the most recent density estimate of the other conditioning variables, until convergence of the ELBO. Computing the Gaussian expectation $\mathbbm{E}_{q(\bbeta)}(\bbeta)$ poses no computational difficulties, whereas, recalling Sections~\ref{sec_3.2} and \ref{sec_4.1}, evaluating the mean $\mathbbm{E}_{q(\bar{\bz}_c)}(\bar{\bz}_c)$, for $c=1, \ldots, C$ of the truncated normals may be challenging when $n_c$ is large. Nonetheless, $n_c$ is typically equal to $1$ or to a small value when factorizing $q(\bar{\bz})$ consistent with the diagonal block structures of ${\bSigma}_{\pst}$ that are implied by most of the models in Sections~\ref{sec_2.1}--\ref{sec_2.4}. This means that the  MF-VB solutions for the local variables $\bar{\bz}$ correspond to tractable low-dimensional truncated normals whose expectation can be computed via efficient routines, such as the one in the \texttt{R} library \texttt{MomTrunc} \citep{galarza2021moments}.

Although MF-VB provides a scalable and widely-applicable solution under the regression models considered in this article, as shown by  \citet{fasano2019} in the context of probit regression with Gaussian priors, the resulting Gaussian approximation $q^{*}_{\textsc{mf}}(\bbeta)$ is characterized by low accuracy, both theoretically and empirically, in high dimensions, especially when $\bar{p}>\bar{n}+\bar{n}_0$. These drawbacks are evident not only in a general underestimation of posterior uncertainty, but also in the tendency to over-shrink the locations and to induce bias in the predictive probabilities, thereby affecting the reliability of Bayesian inference under $q^{*}_{\textsc{mf}}(\bbeta)$. To address these fundamental issues and improve the accuracy of VB in high dimension,  \citet{fasano2019} and \citet{fasano2020} propose a partially-factorized  MF-VB solution (PFM-VB) which replaces the classical mean-field family $\mathcal{Q}_{\textsc{mf}}=\{ q(\bbeta, \bar{\bz}): q(\bbeta, \bar{\bz})=q(\bbeta) \prod_{c=1}^{C} q(\bar{\bz}_c)\}$ with the more flexible partially-factorized one \smash{$\mathcal{Q}_{\textsc{pfm}}=\{ q(\bbeta, \bar{\bz}): q(\bbeta, \bar{\bz})=q(\bbeta \mid \bar{\bz}) \prod_{c=1}^{C} q(\bar{\bz}_c)\}$}, that avoids assuming independence between $\bbeta$ and $\bar{\bz}$ as in mean-field, and only factorizes $q(\bar{\bz})$ as $\prod_{c=1}^{C} q(\bar{\bz}_c)$. The structure of this enlarged family is directly motivated by the form of the actual joint posterior $p(\bbeta, \bar{\bz} \mid \by)$. In fact, as highlighted in Section~\ref{sec_4.2}, $p(\bbeta, \bar{\bz} \mid \by)$ can be re-written as $p(\bbeta \mid \by, \bar{\bz})p(\bar{\bz} \mid \by)$, where $p(\bbeta \mid \by, \bar{\bz})$ is the density of the Gaussian full-conditional in \eqref{eq17}, whereas $p(\bar{\bz} \mid \by)$ is the one of the $(\bar{n}+\bar{n}_0)$-variate truncated normal with full covariance matrix in \eqref{eq18}; see also \citet{holmes_2006}. Therefore, since the Gaussian form of $p(\bbeta \mid \by, \bar{\bz})$ does not seem to pose computational difficulties,  it is reasonable to preserve dependence between $\bbeta$ and $\bar{\bz}$  in $\mathcal{Q}_{\textsc{pfm}}$ and only approximate the intractable multivariate truncated normal density $p(\bar{\bz} \mid \by)$ via the product \smash{$ \prod_{c=1}^{C} q(\bar{\bz}_c)$} of low-dimensional tractable ones. In addition, when the block partitions under MF-VB and PFM-VB coincide, $\mathcal{Q}_{\textsc{mf}} \subset \mathcal{Q}_{\textsc{pfm}}$. Hence, it is guaranteed that the optimum $q_{\textsc{pfm}}^{*}(\bbeta, \bar{\bz})$ under  $\mathcal{Q}_{\textsc{pfm}}$ is never less accurate than $q_{\textsc{mf}}^{*}(\bbeta, \bar{\bz})$, i.e., $\mbox{KL}[q_{\textsc{pfm}}^{*}(\bbeta, \bar{\bz}) || p(\bbeta, \bar{\bz} \mid \by)] \leq \mbox{KL}[q_{\textsc{mf}}^{*}(\bbeta, \bar{\bz}) || p(\bbeta, \bar{\bz} \mid \by)]$. 

The improved accuracy of the PFM-VB approximation, combined with the simple solution of the optimization problem even under the enlarged family $\mathcal{Q}_{\textsc{pfm}}$, have motivated extensions of the original idea in \citet{fasano2019} to multinomial probit \citep{fasano2020} and  Gaussian processes \citep{cao2020scalable}, which can be, in fact, generalized to the whole class of models and priors in Sections~\ref{sec_2}--\ref{sec_3}. To clarify this result,  notice that by the chain rule of the KL divergence it follows that $\mbox{KL}[q(\bbeta \mid \bar{\bz}) \prod_{c=1}^{C} q(\bar{\bz}_c) || p(\bbeta, \bar{\bz} \mid \by)]=\mathbb{E}_{q(\bar{\bz})}(\mbox{KL}[q(\bbeta \mid \bar{\bz}) || p(\bbeta \mid \bar{\bz}, \by)])+\mbox{KL}[\prod_{c=1}^{C} q(\bar{\bz}_c) || p(\bar{\bz} \mid \by)]$, where the first non-negative summand is equal to zero only when $q(\bbeta \mid \bar{\bz})=p(\bbeta \mid \bar{\bz}, \by)$. Therefore, $q_{\textsc{pfm}}^*(\bbeta \mid \bar{\bz})$ is the density of the exact Gaussian full-conditional  in \eqref{eq17}, while the minimizer of $\mbox{KL}[\prod_{c=1}^{C} q(\bar{\bz}_c) || p(\bar{\bz} \mid \by)]$ can be readily obtained by applying the closure under conditioning properties  \citep{Horrace2005SomeRO} of the multivariate truncated normal in \eqref{eq18} to the CAVI equations \smash{$q^{(t)}_{\textsc{pfm}}( \bar{\bz}_c) \propto \exp \{ \mathbbm{E}_{q(\bar{\bz}_{-c})}[\log p(\bar{\bz}_c \mid \by, \bar{\bz}_{-c})] \}$}, for $c=1, \ldots, C$. These results yield a scheme for obtaining $q_{\textsc{pfm}}^{*}(\bbeta, \bar{\bz})$, that is as tractable as the one of MF-VB  in \eqref{eq20}. Specifically, let $\bar{\boeta}^*_{\pst}=\bar{\bz}- \boeta_{\pst}$ and $\bGamma_{\pst(c)}=\bGamma_{\pst[c,c]}- \bGamma_{\pst[c,-c]}(\bGamma_{\pst[-c,-c]})^{-1}\bGamma_{\pst[-c,c]}$. Then, the CAVI equations for  PFM-VB are
\begin{eqnarray}
\begin{split}
&q^*_{\textsc{pfm}}(\bbeta \mid \bar{\bz}) =  \phi_{\bar{p}}(\bbeta-\bV_{\pst}(\bX^{\intercal}_{\pst} \bSigma^{-1}_{\pst}\bar{\boeta}^*_{\pst}+{\bOmega}^{-1}_{\pst} \bxi_{\pst}); \bV_{\pst}), \\
&q^{(t)}_{\textsc{pfm}}(\bar{\bz}_c)\propto \phi_{{n}_c}(\bar{\bz}_c- 
 \mathbbm{E}_{q(\bar{\bz}_{-c})}(\bar{\bmu}_c); \bGamma_{\pst(c)}) \mathbbm{1}(\bar{\bz}_c > {\bf 0}),\\
& \qquad  \qquad  \qquad \qquad  \qquad  \qquad  \qquad  \ \ \   \mbox{for} \  c=1, \ldots, C,
    \end{split}
    \label{eq21}
\end{eqnarray}
where  $ \mathbbm{E}_{q(\bar{\bz}_{-c})}(\bar{\bmu}_c)$ in \eqref{eq21} is defined as $ \mathbbm{E}_{q(\bar{\bz}_{-c})}(\bar{\bmu}_c)=\smash{\bgamma_{\pst[c]}} +\bGamma_{\pst[c,-c]}(\bGamma_{\pst[-c,-c]})^{-1}( \mathbbm{E}_{q(\bar{\bz}_{-c})}(\bar{\bz}_{-c})-\bgamma_{\pst[-c]})$, with the expectation taken with respect to the most recent density estimate of the conditioning variables, whereas the indexing of sub-vectors and matrix blocks is the same as the one detailed in Equation \eqref{eq20}. 

As for the  MF-VB scheme in \eqref{eq20}, also the CAVI for  PFM-VB simply requires to update the mean vectors until convergence of the ELBO. However, unlike for  \eqref{eq20}, such a scheme is only required for the truncated normal components, whereas the solution  $q^*_{\textsc{pfm}}(\bbeta \mid \bar{\bz})$ is already known to coincide with $p(\bbeta \mid \bar{\bz}, \by)$. This gain comes at the cost that, unlike for MF-VB, the approximation $q_{\textsc{pfm}}^{*}(\bbeta)$ of interest  is not available as a direct output of  \eqref{eq21}. Recalling, \citet{fasano2019} and \citet{fasano2020}, this apparent drawback can be easily addressed after noticing that, by \eqref{eq21} and $\bar{\boeta}^*_{\pst}=\bar{\bz}- \boeta_{\pst}$, $q^*_{\textsc{pfm}}(\bbeta)$ is the density of the random variable distributed as a linear combination between a Gaussian, with mean $\bV_{\pst}(-\bX^{\intercal}_{\pst} \bSigma^{-1}_{\pst} \boeta_{\pst}+{\bOmega}^{-1}_{\pst} \bxi_{\pst})$ and covariance matrix $\bV_{\pst}$, and a random vector $\bar{\bz}$ whose joint density is approximated via the product of low-dimensional truncated normals under the CAVI updates in  \eqref{eq21}. Recalling  \eqref{eq14.add}, this construction coincides with the additive representation of a $\mbox{SUN}_{\bar{p},\bar{n}+\bar{n}_0}$  variable that, unlike for the exact SUN posterior in Theorem~\ref{teo1}, relies on a block-diagonal matrix $\bGamma_{\textsc{pfm}}$ with $C$ low-dimensional $n_c \times n_c$ blocks, for $c=1, \ldots, C$. This means that the computational challenges for closed-form inference under the exact SUN posterior discussed in Sections~\ref{sec_3.2} and \ref{sec_4.1} are no more present for the optimal SUN approximating density $q_{\textsc{pfm}}^{*}(\bbeta)$, since the  $(\bar{n}+\bar{n}_0)$-variate Gaussian cumulative distribution functions and truncated normals in Equations~\eqref{eq12}--\eqref{eq16} now factorize as $C$ low-dimensional components that can be effectively evaluated whenever $n_1, \ldots, n_C$ are small-to-moderate. For example,
\begin{eqnarray}
\begin{split}
\mathbb{E}_{\textsc{pfm}}(\bbeta)&=\bV_{\pst}(\bX^{\intercal}_{\pst} \bSigma^{-1}_{\pst}(\mathbb{E}_{\textsc{pfm}}(\bar{\bz})- \boeta_{\pst})+{\bOmega}^{-1}_{\pst} \bxi_{\pst}), \\
\mbox{var}_{\textsc{pfm}}(\bbeta)&= \bV_{\pst} +\bV_{\pst}\bX^{\intercal}_{\pst} \bSigma^{-1}_{\pst}\mbox{var}_{\textsc{pfm}}(\bar{\bz}) \bSigma^{-1}_{\pst}\bX_{\pst}\bV_{\pst},
    \end{split}
    \label{eq22}
\end{eqnarray}
where $\mathbb{E}_{\textsc{pfm}}(\bar{\bz})=[\mathbb{E}_{\textsc{pfm}}(\bar{\bz}^{\intercal}_1), \ldots, \mathbb{E}_{\textsc{pfm}}(\bar{\bz}^{\intercal}_C)]^{\intercal}$ comprises the expectation of each low-dimensional sub-vector $\bar{\bz}_c$, $c=1, \ldots, C$ with respect to its optimal truncated normal approximating density, while $\mbox{var}_{\textsc{pfm}}(\bar{\bz}) $ is a block-diagonal covariance matrix with generic block $\mbox{var}_{\textsc{pfm}}(\bar{\bz})_{[c,c]}$ denoting the covariance matrix of $\bar{\bz}_c$ according to its optimal truncated normal approximation. As previously mentioned, each of these quantities can be effectively evaluated in small-to-moderate dimensions via, e.g., the \texttt{R} library \texttt{MomTrunc} \citep{galarza2021moments}. Recalling \citet{fasano2019}, the computational complexity of PFM-VB is the same as the one for  MF-VB, although the new partially-factorized solution yields  improved accuracy both in theory and in practice. For instance, the authors prove that, unlike for MF-VB, the KL divergence between the PFM-VB approximation  and the exact posterior  goes to $0$ as $\bar{p} \rightarrow \infty$ for any fixed sample size, thereby providing accurate inference in high-dimensional settings at a much lower computational cost than the exact solution.

\subsubsection{Expectation-propagation (EP)}  \label{sec_4.3.2}

EP \citep{Minka2001} provides another well-established procedure for constructing a global approximation $q^*_{\textsc{ep}}(\bbeta)$ of the posterior distribution $p(\bbeta \mid \by)$ \citep[see e.g.,][]{chopin_2017,riihimki2012nested,VehtariJMLR:v21:18-817}, which often yields improved accuracy in practice, relative to VB. Contrarily to the mean-field VB methods presented in Section~\ref{sec_4.3.1} --- which only impose factorized structures for the approximating densities without necessarily assuming a functional form --- EP postulates that the target posterior density itself can be written as a product of factors, also referred to as sites, and then iteratively  approximates each one with an element of a given family of distributions, typically Gaussian for continuous variables or multinomial for discrete ones.
Moreover, in the EP scheme each update is driven by the minimization of a suitable reverse KL, instead of the forward KL as in VB.
This operation tends to improve accuracy \citep[e.g.,][]{chopin_2017} and becomes particularly convenient when the approximating density $q_{\textsc{ep}}(\bbeta)$ belongs to the exponential family, since it simply requires suitable moment matching strategies between $q_{\textsc{ep}}(\bbeta)$ and $p(\bbeta \mid \by)$ \citep[see e.g.,][Chapter 10]{VehtariJMLR:v21:18-817,bishop2006pattern}.

Current implementations of EP for probit  \citep{chopin_2017} and multinomial probit \citep{riihimki2012nested} suggest that these strategies may yield practical gains for the whole class of models in Section~\ref{sec_2}, thus motivating the development of a broadly-applicable unified EP scheme, that is unavailable to date. This section aims at covering such a gap, while providing novel closed-form expressions for moment matching of Gaussian sites leveraging the SUN conjugacy  in Section~\ref{sec_3}, which also yields additional supporting arguments on the accuracy of EP for the models in Section~\ref{sec_2}.

To address such a goal, first notice that, although the likelihood in \eqref{eq1} is  general, all the relevant examples discussed in Section~\ref{sec_2} admit a factorized form $\prod_{c=1}^{C} \Phi_{\bar{n}_c}(\bar{\by}_{0[c]}+\bar{\bX}_{0[c]}\bbeta; \bar{\bSigma}_{0[c,c]})$ for the intractable quantity $\Phi_{\bar{n}_0}(\bar{\by}_{0}+\bar{\bX}_{0}\bbeta; \bar{\bSigma}_{0})$, where $\bar{\by}_0=(\bar{\by}^{\intercal}_{0[1]}, \ldots, \bar{\by}^{\intercal}_{0[C]})^{\intercal}$, $\bar{\bX}_0=(\bar{\bX}^{\intercal}_{0[1]}, \ldots, \bar{\bX}^{\intercal}_{0[C]})^{\intercal}$ and $\bar{\bSigma}_{0}$ is a block-diagonal matrix with generic block \smash{$\bar{\bSigma}_{0[c,c]}$}, for every $c=1, \ldots, C$. As discussed in Sections~\ref{sec_4.2} and \ref{sec_4.3.1}, this factorization is implied by the conditional independence among the latent utilities, which yields to tractable one-dimensional (e.g., probit and tobit) or low-dimensional (e.g., multinomial probit) factors $\Phi_{\bar{n}_c}(\bar{\by}_{0[c]}+\bar{\bX}_{0[c]}\bbeta; \bar{\bSigma}_{0[c,c]})$. Therefore, under these models, the likelihood in \eqref{eq1} is equal to
\begin{equation}
\begin{split}
&p(\by \mid \bbeta)=p(\bar{\by}_1 \mid \bbeta)p(\bar{\by}_0 \mid \bbeta)\\
& \textstyle \propto \phi_{\bar{n}_1}(\bar{\by}_1{-}\bar{\bX}_1 \bbeta;\bar{\bSigma}_1){\prod\nolimits_{c=1}^{C}} \Phi_{\bar{n}_c}(\bar{\by}_{0[c]}{+}\bar{\bX}_{0[c]}\bbeta; \bar{\bSigma}_{0[c,c]}),
\end{split}
    \label{eq23}
\end{equation}
thus providing a general factorized structure that motivates  EP. For ease of notation and presentation, this routine is first derived below under the Gaussian prior $p(\bbeta) = \phi_{\bar{p}} (\bbeta - \bxi; \bOmega)$, and  subsequently extended to the general class of  SUN prior distributions. Updating $p(\bbeta) = \phi_{\bar{p}} (\bbeta - \bxi; \bOmega)$ with the likelihood in \eqref{eq23} yields the posterior  $p(\bbeta \mid \by)$ which can be more conveniently re-expressed as
\begin{equation}
\begin{split}
 &   p(\bbeta \mid \by) \propto p(\bbeta \mid \bar{\by}_1 ) p(\bar{\by}_0 \mid \bbeta)
 \\
& \textstyle \propto   \phi_{\bar{p}} (\bbeta - \bxi_{\pst}; \bOmega_{\pst}) {\prod\nolimits_{c=1}^{C}} \Phi_{\bar{n}_c}(\bar{\by}_{0[c]}{+}\bar{\bX}_{0[c]}\bbeta; \bar{\bSigma}_{0[c,c]}) \\
&\textstyle \quad =l_0(\bbeta)\prod\nolimits_{c=1}^{C}l_c(\bbeta)=  \prod\nolimits_{c=0}^{C}l_c(\bbeta),
\end{split}
\label{eq24}
\end{equation}
where $l_c(\bbeta)=\Phi_{\bar{n}_c}(\bar{\by}_{0[c]}+\bar{\bX}_{0[c]}\bbeta; \bar{\bSigma}_{0[c,c]})$, $c=1, \ldots, C$, correspond to the Gaussian cumulative distribution function terms in likelihood \eqref{eq23}, whereas $l_0(\bbeta)=p(\bbeta \mid \bar{\by}_1)=\phi_{\bar{p}} (\bbeta - \bxi_{\pst}; \bOmega_{\pst})$ is the conditional density obtained by updating the Gaussian prior $ \phi_{\bar{p}} (\bbeta - \bxi; \bOmega)$ for $\bbeta$ with the tractable factor $\phi_{\bar{n}_1}(\bar{\by}_1-\bar{\bX}_1 \bbeta;\bar{\bSigma}_1)$ in likelihood \eqref{eq23}. As a direct consequence of the results in Section~\ref{sec_3.2}, this conditional density can be obtained in closed form and coincides with the one of a Gaussian $\mbox{N}_{\bar{p}} (\bxi_{\pst}, \bOmega_{\pst})$ having parameters defined as in Theorem~\ref{teo1}. Such a density acts as an intermediate prior in \eqref{eq24} to be updated with the intractable likelihood terms for obtaining the posterior $  p(\bbeta \mid \by) $.

Recalling, for instance,  \citet{VehtariJMLR:v21:18-817}, EP approximates the above posterior with a density $q_{\textsc{ep}}(\bbeta)$ that has the same factorized form of $ p(\bbeta \mid \by) $ in \eqref{eq24}, and is made of $C+1$ Gaussian sites. Hence
\begin{equation}
\begin{split}
 \textstyle   q_{\textsc{ep}}(\bbeta)\propto {\prod\nolimits_{c=0}^{C}}q_c(\bbeta)& = \textstyle {\prod\nolimits_{c=0}^{C}}\exp ( {-}0.5\bbeta^\intercal \bQ_{c} \bbeta {+} \bbeta^\intercal \br_{c})\\
    &=\exp ( -0.5\bbeta^\intercal \bQ_{\textsc{ep}} \bbeta + \bbeta^\intercal \br_{\textsc{ep}}),
    \end{split}
\label{eq25}
\end{equation}
where $ \br_{c}$ and $\bQ_{c}$ define the natural parameters associated with the local Gaussian site $c$, for each $c=0, \ldots, C$, whereas $\br_{\textsc{ep}}= \sum_{c=0}^{C}  \br_{c}$ and \smash{$\bQ_{\textsc{ep}}= \sum_{c=0}^{C}  \bQ_{c}$} denote those of the Gaussian EP approximation $ q_{\textsc{ep}}(\bbeta)$ for $ p(\bbeta \mid \by) $. Consistent with the above expressions, the ideal goal of EP would be to obtain the optimal $\br^*_{\textsc{ep}}$ and $\bQ^*_{\textsc{ep}}$ such that the induced Gaussian density $q^*_{\textsc{ep}}(\bbeta)$ under \eqref{eq25}  is as close as possible to the exact $p(\bbeta \mid \by)$ in \eqref{eq24} under the reverse KL divergence $ \mbox{KL}[p(\bbeta \mid \by) || q_{\textsc{ep}}(\bbeta)]$. Recalling \citet[][Chapter 10]{bishop2006pattern}, the solution of this optimization problem relies on a simple moment matching, which implies that  $\br^*_{\textsc{ep}}=[\mbox{var}(\bbeta \mid \by)]^{-1}\mathbb{E}(\bbeta \mid \by)$ and $\bQ^*_{\textsc{ep}}=~[\mbox{var}(\bbeta \mid \by)]^{-1}$, or, alternatively, $\bxi^*_{\textsc{ep}}=\mathbb{E}(\bbeta \mid \by)$ and $\bOmega^*_{\textsc{ep}}=\mbox{var}(\bbeta \mid \by)$, where $\bxi^*_{\textsc{ep}}$ and $\bOmega^*_{\textsc{ep}}$ denote the mean vector and the covariance matrix of the Gaussian EP approximation. As discussed in Section~\ref{sec_3.2}, the exact posterior is a SUN, and, hence, computing the associated moments is computationally challenging in general settings. In fact, such computational bottlenecks are those motivating the approximate schemes  in Section~\ref{sec_4.3}.

To circumvent the aforementioned issue, EP relies on an iterative scheme which progressively~improves \smash{$\br_{\textsc{ep}}=\sum_{c=0}^{C}  \br_{c}$} and \smash{$\bQ_{\textsc{ep}}=\sum_{c=0}^{C}  \bQ_{c}$} by sequentially updating  each term $(\br_{c}, \bQ_{c})$, for $c=1, \ldots, C$, keeping fixed the others at their previous estimate \citep[see e.g.,][]{VehtariJMLR:v21:18-817}. Let \smash{$l_{(-c)}(\bbeta){=}\prod_{c' \neq c}l_{c'}(\bbeta)$} and \smash{$q_{(-c)}(\bbeta)=\prod_{c' \neq c}q_{c'}(\bbeta)$} denote the product among the factors in~\eqref{eq24}--\eqref{eq25}, respectively, excluding the $c$--th one. Then, EP proceeds by optimizing, for every site $c$, a more tractable approximation for the reverse KL divergence in which the exact posterior $p(\bbeta \mid \by) \propto l_{(-c)}(\bbeta)l_c(\bbeta)$ in~\eqref{eq24} is replaced by the intermediate hybrid density defined as \smash{$p^{(t_c)}(\bbeta \mid \by) \propto q^{(t_c)}_{(-c)}(\bbeta)l_c(\bbeta)$}, where $t_c$ is the step of the algorithm which updates the site $c$ at the $t$--th iteration. Employing \smash{$q^{(t_c)}_{(-c)}(\bbeta)$} instead of $l_{(-c)} (\bbeta)$ yields a more tractable density since, by Equation \eqref{eq25}, \smash{$q^{(t_c)}_{(-c)}(\bbeta)$} is the kernel of a multivariate Gaussian  with natural parameters \smash{$\br^{(t_c)}_{\textsc{ep}(-c)}$} and \smash{$\bQ^{(t_c)}_{\textsc{ep}(-c)}$} corresponding to $\br_{\textsc{ep}}- \br_{c}$ and $\bQ_{\textsc{ep}}- \bQ_{c}$, respectively, when $\br_{\textsc{ep}}$, $\br_{c}$, $\bQ_{\textsc{ep}}$ and $\bQ_{c}$ are fixed at their most recent estimate. Therefore,  $p^{(t_c)}(\bbeta \mid \by)$ has a single Gaussian cumulative distribution function term \smash{$l_c(\bbeta)=\Phi_{\bar{n}_c}(\bar{\by}_{0[c]}{+}\bar{\bX}_{0[c]}\bbeta; \bar{\bSigma}_{0[c,c]})$}.  Adapting the results in Section~\ref{sec_3}, this yields the hybrid density $p^{(t_c)}(\bbeta \mid \by) \propto \smash{q^{(t_c)}_{(-c)}(\bbeta)l_c(\bbeta)}$ with
\begin{eqnarray*}
\begin{split}
q^{(t_c)}_{(-c)}(\bbeta)&= \phi_{\bar{p}}(\bbeta-(\bQ^{(t_c)}_{\textsc{ep}(-c)})^{-1}\br^{(t_c)}_{\textsc{ep}(-c)}; (\bQ^{(t_c)}_{\textsc{ep}(-c)})^{-1}),\\
l_c(\bbeta)&= \Phi_{\bar{n}_c}(\bar{\by}_{0[c]}+\bar{\bX}_{0[c]}\bbeta; \bar{\bSigma}_{0[c,c]}),
\end{split}
\end{eqnarray*} 
which implies that $p^{(t_c)}(\bbeta \mid \by)$ coincides with the density of the \smash{$\mbox{SUN}_{\bar{p},\bar{n}_c}(\bxi_c,\bOmega_c,\bDelta_c,\bgamma_c,\bGamma_c)$} with
\begin{eqnarray*}
\begin{split}
\bxi_c&=(\bQ^{(t_c)}_{\textsc{ep}(-c)})^{-1}\br^{(t_c)}_{\textsc{ep}(-c)},  \quad \bOmega_c=(\bQ^{(t_c)}_{\textsc{ep}(-c)})^{-1},\\
\bDelta_c&= \bar{\bOmega}_c\bomega_c\bar{\bX}^{\intercal}_{0[c]}\bs_c^{-1}, \qquad \ \ \ \bgamma_c=\bs_{c}^{-1}(\bar{\by}_{0[c]}+\bar{\bX}_{0[c]}\bxi_c),\\
\bGamma_c&=\bs_{c}^{-1}(\bar{\bSigma}_{0[c,c]}+\bar{\bX}_{0[c]} \bOmega_c \bar{\bX}^{\intercal}_{0[c]})\bs_{c}^{-1},
\end{split}
\end{eqnarray*} 
where $\bs_{c}=[(\bar{\bSigma}_{0[c,c]}+\bar{\bX}_{0[c]} \bOmega_c \bar{\bX}^{\intercal}_{0[c]}) \odot \bI_{\bar{n}_c}]^{1/2}$. Therefore, unlike for the exact SUN posterior, this hybrid SUN is much more tractable since the dimension of the cumulative distribution function term is $\bar{n}_c$, and not \smash{$\sum_{c=1}^C \bar{n}_c$} as in $p(\bbeta \mid \by)$. In fact, as previously discussed, $\bar{n}_c$ is either equal to $1$ or to a low value under most of the models outlined in Sections~\ref{sec_2.1}--\ref{sec_2.4}. This means that inference under the SUN with density \smash{$p^{(t_c)}(\bbeta \mid \by)$} can be performed via the closed-form expressions in Section~\ref{sec_3.2}, which can be effectively evaluated when $\bar{n}_c$ is small; see also Section~\ref{sec_4.1}. In particular, it is possible to compute the expectation \smash{$\mathbb{E}^{(t_c)}(\bbeta \mid \by)$} and variance \smash{$\mbox{var}^{(t_c)}(\bbeta \mid \by)$} of $\bbeta$ with respect to the hybrid density \smash{$p^{(t_c)}(\bbeta \mid \by)$} via expressions \eqref{eq13} evaluated at the parameters $\bxi_c,\bOmega_c,\bDelta_c,\bgamma_c$ and $\bGamma_c$. Alternatively, leveraging the additive representation of the SUN in \eqref{eq14.add}, it follows that
\begin{eqnarray}
\begin{split}
\mathbb{E}^{(t_c)}(\bbeta \mid \by) &= \bxi_{c}+\bomega_{c}\bDelta_{c}\bGamma^{-1}_{c}\mathbb{E}(\bU_{1c})=\bxi^{(t_c)}_{\textsc{ep}}, \\
\mbox{var}^{(t_c)}(\bbeta \mid \by) &= \bOmega_{c}-\bomega_{c}\bDelta_{c}\bGamma^{-1}_{c}\bDelta^{\intercal}_{c}\bomega_{c}\\
& \quad +\bomega_{c}\bDelta_{c}\bGamma^{-1}_{c}\mbox{var}(\bU_{1c}) \bGamma^{-1}_{c}\bDelta^{\intercal}_{c}\bomega_{c}=\bOmega^{(t_c)}_{\textsc{ep}},
    \end{split}
    \label{eq26}
\end{eqnarray}
where  $\bU_{1c} \sim \mbox{TN}_{\bar{n}_c}(-\bgamma_{c}; {\bf0}, \bGamma_{c})$ is a low-dimensional truncated normal whose expectation $\mathbb{E}(\bU_{1c})$ and variance $\mbox{var}(\bU_{1c})$ can be effectively computed via  \texttt{R} library \texttt{MomTrunc} \citep{galarza2021moments}, due to the small value of $\bar{n}_c$. This implies that the reverse KL can be easily optimized via moment matching when $p(\bbeta \mid \by)$ is replaced by \smash{$p^{(t_c)}(\bbeta \mid \by)$}, thereby obtaining the updated estimates \smash{$\br^{(t_c)}_{\textsc{ep}}$} and \smash{$\bQ^{(t_c)}_{\textsc{ep}}$}  for the parameters of interest $\br_{\textsc{ep}}$ and \smash{$\bQ_{\textsc{ep}}$} at step $t_c$, defined as
\begin{eqnarray*}
\begin{split}
\br^{(t_c)}_{\textsc{ep}} &=(\bOmega^{(t_c)}_{\textsc{ep}})^{-1}\bxi^{(t_c)}_{\textsc{ep}}, \qquad
\bQ^{(t_c)}_{\textsc{ep}} &=(\bOmega^{(t_c)}_{\textsc{ep}})^{-1}.
    \end{split}
\end{eqnarray*}
Concurrently, the updated parameters at site $c$ --- which are required for the subsequent steps --- are \smash{$\br^{(t_c)}_{c}=\br^{(t_c)}_{\textsc{ep}}-\br^{(t_c)}_{\textsc{ep}(-c)}$} and \smash{$\bQ^{(t_c)}_{c}=\bQ^{(t_c)}_{\textsc{ep}}-\bQ^{(t_c)}_{\textsc{ep}(-c)}$}.

The above scheme is iterated multiple times $t \in \{1,2, \ldots\}$ and for each site $c=1, \ldots, C,$ until convergence to a stationary point. Note that in this routine site $c=0$ does not require to be updated sequentially. Recalling  \citet{chopin_2017} and \citet{VehtariJMLR:v21:18-817}, $l_0(\bbeta)$ corresponds to the tractable Gaussian density in \eqref{eq24} and, hence, this term can be analytically matched to $q_0(\bbeta)$  in \eqref{eq25}, obtaining $\br_0=\bOmega^{-1}_{\pst}\bxi_{\pst}$ and $\bQ_0=\bOmega^{-1}_{\pst}$, where $\bxi_{\pst}$ and $\bOmega_{\pst}$ are defined as in Theorem~\ref{teo1}. We shall also emphasize that the aforementioned EP scheme can yield, as a direct by-product, an approximation of the marginal likelihood $p(\by)$. A detailed presentation of the step-by-step procedure to obtain such an estimate can be found in Appendix E of \citet{VehtariJMLR:v21:18-817}, which shows that a key condition to compute such an approximation is the availability of  the normalizing constant for the hybrid density \smash{$p^{(t_c)}(\bbeta \mid \by)$}. Interestingly, this quantity is available in closed form for the EP scheme discussed above since \smash{$p^{(t_c)}(\bbeta \mid \by)$} is the density of a \smash{$\mbox{SUN}_{\bar{p},\bar{n}_c}(\bxi_c,\bOmega_c,\bDelta_c,\bgamma_c,\bGamma_c)$}. Hence, recalling Section~\ref{sec_3}, its normalizing constant is $\Phi_{\bar{n}_c}(\bgamma_c; \bGamma_c)$. Since $\bar{n}_c$ is small, also this quantity can be effectively evaluated using, for example, the \texttt{R} library  \texttt{TruncatedNormal} \citep{Botev_2016}.

Although EP is often more accurate than VB, it shall be noted that state-of-the-art implementations build on weaker theoretical guarantees \citep[e.g.,][]{bishop2006pattern,chopin_2017,VehtariJMLR:v21:18-817} and,  as  discussed in the Supplementary Materials,   are more computationally demanding. In particular, even the new EP implementation  proposed in Section~S3 --- which noticeably reduces  the currently reported per-iteration cost in probit regression from $\mathcal{O}(np^2)=\mathcal{O}(\bar{n}_0\bar{p}^2)$ \citep{chopin_2017} to $\mathcal{O}(np \cdot \mbox{min}\{n,p\})=\mathcal{O}(\bar{n}_0\bar{p}\cdot \mbox{min}\{\bar{n}_0,\bar{p}\})$ --- is still not competitive with the $\mathcal{O}(n \cdot \mbox{min}\{n,p\})=\mathcal{O}(\bar{n}_0\cdot \mbox{min}\{\bar{n}_0,\bar{p}\})$ cost of MF-VB and PFM-VB \citep{fasano2019}. Moreover, there is no guarantee that the EP  solution minimizes the global reverse $ \mbox{KL}[p(\bbeta \mid \by) || q_{\textsc{ep}}(\bbeta)]$, nor that the routine always converges in general. Nonetheless, empirical evidence typically reports remarkable EP accuracy, which is also confirmed by the simulations in Section~\ref{sec_5}. Recalling \citet[][]{bishop2006pattern} an intuition for this notable performance is that, at each EP iteration, the sites are updated to be most accurate in regions of high posterior probability.  \citet{DehaeneBarthelme2018} provide more formal arguments,  which show that in asymptotic settings the discrepancy between the EP solution and the exact posterior goes to $0$ faster than, for instance, Laplace approximation. These results are intimately related to the log-concavity of the target posterior. Interestingly, as shown in \citet[][Section 3.1]{arellano2021}, SUNs are log-concave. Hence, the results in Section~\ref{sec_3} also provide further support to EP under the models  in Section~\ref{sec_2}.

Before concluding the analysis of EP, notice that extending the above derivations to the case of a more general SUN prior poses no conceptual difficulties. In fact, the Gaussian density and distribution function appearing in the prior in~\eqref{eq11} can be disentangled and treated as two separate sites appearing in the factorized target. In particular, the first exact site $l_0(\bbeta)= \phi_{\bar{p}} (\bbeta - \bxi_{\pst}; \bOmega_{\pst})$ remains unchanged, as it still arises from the combination of the Gaussian density $ \phi_{\bar{p}} (\bbeta - \bxi; \bOmega)$ in the prior and the likelihood term $\phi_{\bar{n}_1}(\bar{\by}_1-\bar{\bX}_1 \bbeta;\bar{\bSigma}_1)$. Conversely, the distribution function term in the prior can be simply addressed by adding a site $l_{C+1}(\bbeta) = \Phi_{\bar{n}}(\bgamma + \bDelta^\intercal \bar{\bOmega}^{-1} \bomega^{-1}(\bbeta-\bxi); \bGamma - \bDelta^\intercal \bar{\bOmega}^{-1} \bDelta)$, to be approximated via an extra term $q_{C+1}(\bbeta)$ in~\eqref{eq25}. As such, the only hindrances might arise from the computation of the moments for the hybrid distribution with kernel \smash{$l_{C+1}(\bbeta) \prod\nolimits_{c=0}^{C}q^{(t_{C+1})}_c(\bbeta)$}, which still corresponds to a SUN of dimensions $\bar{p}$ and $\bar{n}$. However, as mentioned in Section~\ref{sec_3.1}, the SUN prior often relies on a low $\bar{n}$, or alternatively factorizes as the product of independent  skew-normals. In the former case, computations in low dimensions remain feasible, while in the latter case the $(C+1)$-th exact site can be further disentangled as the product of $\bar{p}$ sites, each one involving a tractable univariate Gaussian cumulative distribution function. This flexibility in choosing the factorization of the target distribution is a general characteristic of EP. The two extreme cases correspond, respectively, to considering the target as one single site or decomposing it into the finest factorization allowed by its analytical formulation. Any intermediate situation  leads to a valid EP routine, as described above, possibly with a trade-off between accuracy of the resulting approximation and complexity of the required computations. Finer factorizations lead to simpler moment matching calculations. Conversely, coarser factorizations yield more accurate, but expensive, approximations.

\section{Empirical studies}\label{sec_5}

\begin{table*}[b!]
\renewcommand{\arraystretch}{1}
\centering
\caption{\footnotesize{Runtimes, in seconds, of two  strategies to sample $5000$ realizations from the posterior distribution in tobit regression with $n=200$. (NUTS): \texttt{rstan} implementation of No-U-Turn HMC sampler. (i.i.d.): i.i.d.\ sampling from the exact SUN posterior via Equation \eqref{eq14.add} leveraging the \texttt{R} library \texttt{TruncatedNormal}.}}
\label{table_runtimes_MC}
\begin{tabular}[c]{ccllllllll}
 \multicolumn{2}{c}{}  &  \multicolumn{8}{c}{\textit{p}}   \\
\midrule
\textsc{Censoring} & \textsc{Method} \qquad \qquad  & $10$ & $20$ & $50$ & $100$ & $200$ & $400$ & $800$ & $1200$ \\ 
   \midrule
\multirow{2}{*}{$\kappa=0.85$}
    & NUTS \qquad \qquad & $2.32$ & $4.61$ & $9.87$ & $26.17$ & $50.21$ & $134.95$ & $323.73$ & $659.25$ \\
    & i.i.d.\ \qquad \qquad & $2.94$ & $2.85$ & $18.97$ & $15.98$ & $7.71$ & $3.55$ & $4.94$ & $9.26$ \\
\midrule   
\multirow{2}{*}{$\kappa=0.50$}
    & NUTS \qquad \qquad& $1.62$ & $2.04$ & $3.78$ & $10.90$ & $39.06$ & $110.73$ & $546.07$ & $2128.08$  \\
    & i.i.d.\ \qquad \qquad& $1.30$ & $1.21$ & $1.20$ & $1.84$ & $2.58$ & $2.33$ & $4.21$ & $8.25$  \\
\midrule 
\multirow{2}{*}{$\kappa=0.15$}
    & NUTS \qquad \qquad& $0.91$ & $1.18$ & $2.27$ & $4.60$ & $20.08$ & $87.10$ & $619.58$ & $1973.92$  \\
    & i.i.d.\ \qquad \qquad & $0.13$ & $0.10$ & $0.14$ & $0.23$ & $0.56$ & $1.13$ & $2.91$ & $6.91$ \\
\hline
\end{tabular}
\end{table*}

Insightful empirical assessments of the  methods in Sections~\ref{sec_3}--\ref{sec_4}, under selected regression models, can be found in  \citet{chopin_2017,durante_2019,fasano2020,cao2020scalable,fasano2019,fasano2021closed} and \citet{benavoli2020skew,benavoli2021unified}; refer also to the GitHub repositories \texttt{ProbitSUN}, \texttt{PredProbitGP}, \texttt{Probit-PFMVB} and \texttt{Dynamic-Probit-PFMVB}. These studies encompass analyses of probit regression, multinomial probit, dynamic probit, probit Gaussian processes, skewed Gaussian processes and possible combinations of these constructions, but do not cover tobit regression for which SUN conjugacy has been established  in the present article and, hence, the practical consequences of this result and the associated computational methods remain unexplored.

To address such a gap, we provide empirical evidence for the performance of the computational methods  in Section~\ref{sec_4}, focusing on standard tobit regression as in \eqref{eq9}. In accomplishing this goal, we simulate a total of $n=n_0+n_1=200$ observations from a tobit model, under three different proportions of censored observations $\kappa=n_0/n \in \{0.15,0.50,0.85\}$. This choice allows to cover a broad spectrum of scenarios which ranges from a model more similar to a Gaussian linear regression, when $\kappa=0.15$, to one closely mimicking an unbalanced probit regression, when $\kappa=0.85$. The $p$ unit-specific predictors in $\bx_i$, $i=1, \ldots, n$, are instead simulated from standard Gaussians, except for the intercept term, whereas the regression coefficients in $\bbeta$ are generated from a uniform distribution in the range $[-5,5]$. Exploiting the latent utility interpretation of tobit regression, the responses $y_i$, $i=1, \ldots, n$ are obtained by first simulating the associated  utilities $z_i$, $i=1, \ldots, n$ from the $\mbox{N}(\bx^{\intercal}_i\bbeta, 1)$, and then setting $y_i=z_i \mathbbm{1}(z_i>z_{\textsc{t}})$ for every $i=1, \ldots, n$, where $z_{\textsc{t}}$ is a pre-specified threshold to obtain the desired proportion of censored observations under the three different settings of $\kappa$. Recalling Section~S1.2 in the Supplementary Materials, this  threshold poses no difficulties in Bayesian inference since it directly enters the intercept term. To evaluate accuracy and computational efficiency at varying dimensions, these datasets are simulated for different values of $p \in \{10, 20, 50, 100, 200, 400, 800, 1200 \}$.  Posterior inference under the datasets produced for each combination of $\kappa$ and $p$ relies on  spherical Gaussian priors $\mbox{N}_p({\bf 0},\omega_p^2 \bI_p)$, with $\omega_p^2=25 \cdot 10/p$ inducing increasing shrinkage in high dimensions. In combination with the recommended practice of standardizing the predictors to have  mean $0$ and standard deviation $0.5$ \citep[e.g.,][]{chopin_2017}, such a weakly informative prior is intended to control the overall variance of the linear predictor \citep[see e.g.,][]{fasano2019} so as to constrain it within a sensible range of variation for the models analyzed in this study \citep[e.g.,][]{Gelman2008}, regardless of  $p$. This facilitates comparison across different dimensions.

\begin{figure*}[t]
\centering
    \includegraphics[width=1.03\textwidth]{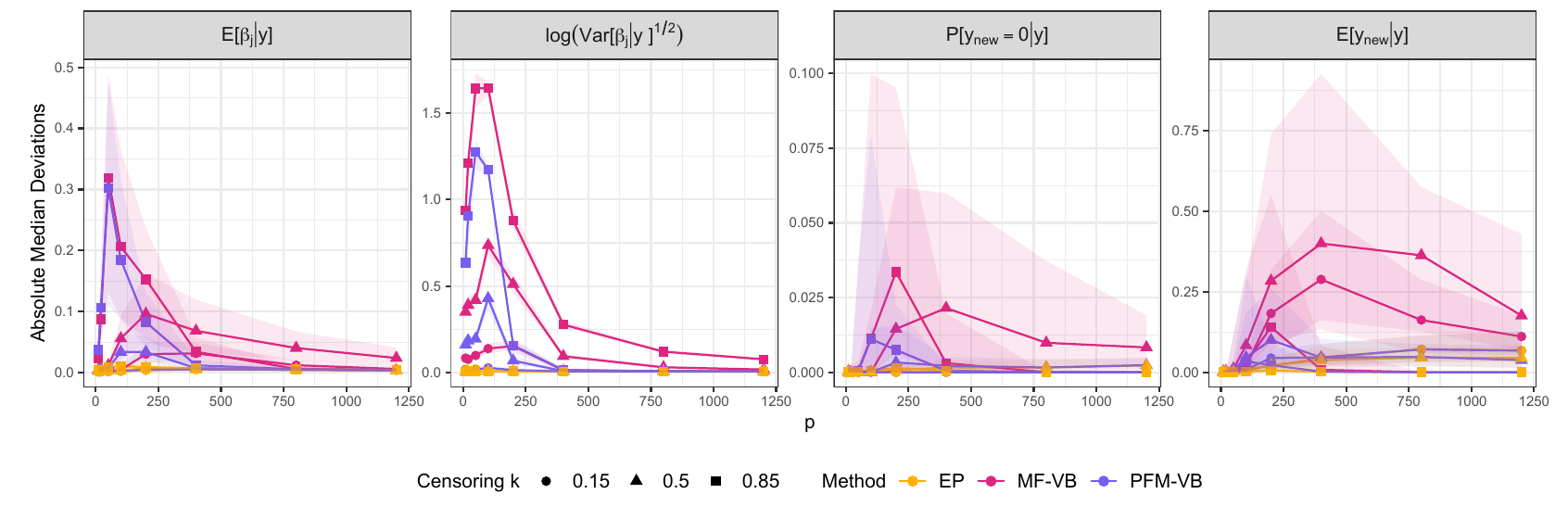}
      \vspace{-18pt}
    \caption{\footnotesize{For four functionals of interest and different settings of $\kappa \in \{0.15,0.50,0.85\}$, trajectories for the median of the absolute differences, at varying $p$, between an accurate Monte Carlo estimate of such functionals via i.i.d.\ sampling from the exact SUN posterior and their approximation provided by mean-field variational Bayes (MF-VB), partially-factorized variational Bayes (PFM-VB) and expectation-propagation (EP) under tobit regression, with $n=200$. The shaded areas correspond to the first and third quartiles computed from the absolute differences. See the online article for the color version of this figure.}
}
    \label{figure:1}
\end{figure*}

Table~\ref{table_runtimes_MC} illustrates the computational gains in sampling-based methods which can be obtained by leveraging routines that exploit directly the SUN conjugacy in Section~\ref{sec_3}. This is accomplished by comparing, for every combination of $\kappa$ and $p$, the runtimes to obtain $5000$ samples from the exact posterior distribution of $\bbeta$ under both the routinely-used \texttt{rstan} implementation of the No-U-Turn HMC algorithm, and the i.i.d.\ sampler which exploits the additive representation of the SUN posterior in  \eqref{eq14.add}; refer to the code at  \url{https://github.com/niccoloanceschi/TobitSUN} and to \citet{chopin_2017} for details on the implementation of the HMC sampler in the class of models analyzed.  The i.i.d.\ scheme leverages instead  the \texttt{R} library \texttt{TruncatedNormal}  \citep{Botev_2016} to sample the multivariate truncated normal component in \eqref{eq14.add}.  As discussed in Sections~\ref{sec_4.1} and \ref{sec_4.2}, such a task is inherently related, in terms of implementation and computational cost, to that of evaluating the Gaussian cumulative distribution functions required to conduct posterior inference under the closed-form expressions in Section~\ref{sec_3.2}. Nonetheless, Monte Carlo inference under  i.i.d.\ samples has the additional benefit of allowing evaluation of any functional, even beyond  those derived in closed form in Section~\ref{sec_3.2}, from a single set of samples, thereby motivating our focus on sampling-based methods which allow for a more comprehensive assessment. 

Consistently with related findings on probit \citep{durante_2019} and multinomial probit \citep{fasano2020}, Table~\ref{table_runtimes_MC} confirms the computational gains of  i.i.d.\ sampling relative to HMC in almost all the settings of $\kappa$ and $p$, especially when $p$ is large. In fact, while high-dimensional regimes are often challenging for HMC, under \eqref{eq14.add} $p$ only controls the dimension of the multivariate Gaussian, which is feasible to sample from, even for a large $p$. As discussed in Sections~\ref{sec_4.1}--\ref{sec_4.2}, more problematic for the  i.i.d.\ scheme is the number of censored data $n_0$, which defines the dimension of the truncated normal in \eqref{eq14.add}. This issue can be clearly seen in the increments of the runtimes under  i.i.d.\  sampling when the  censoring percentage  grows from $15\%$ to $85\%$. Nonetheless, the procedure is still competitive relative to  HMC in these small-to-moderate $n_0$ settings. Notice also an increment in the runtime for the setting $\kappa=0.85$ (i.e., $n_0=170$), when $p \approx n_0/2$. In such a regime, the method by \citet{Botev_2016} experiences low acceptance probabilities with a trend over $n_0$ that  is reminiscent of the double-descent phenomenon in high-dimensional regression \citep{hastie_2022}. This  deserves further investigations.

\begin{figure*}[t]
\centering
    \includegraphics[width=0.85\textwidth]{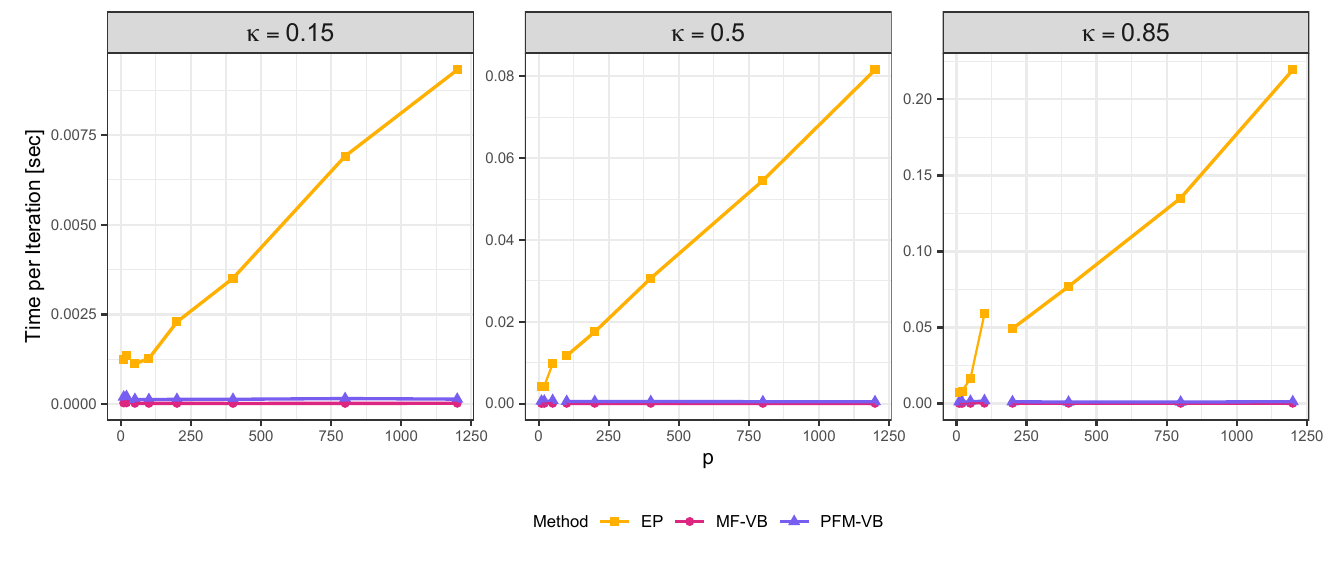}
        \vspace{-12pt}
    \caption{\footnotesize{Runtimes, in seconds, for each iteration of mean-field variational Bayes (MF-VB), partially-factorized variational Bayes (PFM-VB) and expectation-propagation (EP) in tobit regression, for $n=200$ and different values of $p$ and $\kappa$. The runtimes are obtained as the median over 20 repetitions. The line breaks separate the result for two alternative implementations of each routine, optimizing the computational cost in $p$ and $n_0$. For the two variational procedures, the cost per iteration is $\mathcal{O}(n_0 \cdot \min\{n_0,p\})$, while for EP it becomes $\mathcal{O}(n_0 p \cdot \min\{n_0,p\})$. See the online article for the color version of this figure.}
    }
    \label{figure:2}
\end{figure*}

As clarified in Table~\ref{table_runtimes_MC}, the moderate dimensions of the simulated datasets would still allow posterior inference under the closed-form solutions and i.i.d.\ sampling schemes presented in Sections~\ref{sec_4.1} and \ref{sec_4.2}. Nonetheless, as previously discussed,  when $n_0$  grows, these procedures become computationally impractical, thus motivating also the assessment of the more scalable approximate methods presented in Section~\ref{sec_4.3}. The relevant outcomes of these performance comparisons are reported in Figures~\ref{figure:1}--\ref{figure:2} and in Table~\ref{table_Niter}, with a focus on both accuracy and scalability. In particular, Figure~\ref{figure:1} provides insights on the accuracy of MF-VB, PFM-VB and EP in approximating key posterior functionals of interest at varying $p$, and for the three different settings of $\kappa$. These quantities include the posterior mean and variance of each $\beta_j$ for $j=1, \ldots, p$, along with predictive measures for  the expected value of the response $ \mathbb{E}[\phi(\bx^{\intercal}_{\textsc{new},i} \bbeta) + (\bx^{\intercal}_{\textsc{new},i} \bbeta)\Phi(\bx^{\intercal}_{\textsc{new},i} \bbeta)\mid \by]$ and the probability of the censoring event $\mathbb{E}[\Phi(-\bx^{\intercal}_{\textsc{new},i} \bbeta) \mid \by]$, both computed for $200$ test observations whose predictors are simulated as for the original training data. For such functionals, Figure~\ref{figure:1} displays the medians and quartiles of the absolute differences between the corresponding Monte Carlo estimates under i.i.d.\ sampling from the exact SUN posterior and the approximations provided by the three methods analyzed, for varying $\kappa$ and~$p$. In the first two panels, the three quartiles are computed on the $p$ absolute differences associated with coefficients $\beta_1, \ldots, \beta_p$, whereas in the last two panels these summaries are calculated on the $200$ absolute differences for the $200$ test units. For what concerns the initialization of the variational routines, we consider the default setting which sets to zero the elements of the vectors $\mathbbm{E}_{q(\bbeta, \bar{\bz}_{-c})}(\bmu_c)$ and $\mathbbm{E}_{q(\bar{\bz}_{-c})}(\bar{\bmu}_c)$ in Equations~\eqref{eq20} and~\eqref{eq21}, respectively, for all $c=1,\dots,C$. Analogously, in the case of EP we initialize the starting global approximation  to  the Gaussian density $l_0(\bbeta)=p(\bbeta \mid \bar{\by}_1)$, which corresponds to setting to zero the elements of the vectors $\br_{c}$ and matrices $\bQ_{c}$ in Equation~\eqref{eq25}, for all $c=1,\dots,C$. In the absence of additional information, these initializations provide a sensible choice, which proved effective in a large number of experiments.

\begin{table*}[b]
\centering
\renewcommand{\arraystretch}{1}
        \vspace{-12pt}
\caption{\footnotesize{Total runtimes in seconds (pre-computations, total iterations and post-computations),  and number of iterations (within square brackets)  required to reach convergence for mean-field variational Bayes (MF-VB), partially-factorized variational Bayes (PFM-VB) and expectation-propagation (EP) under tobit regression, with $n=200$. The reported runtimes are the medians over 20 repetitions.}}
\label{table_Niter}
\begin{adjustbox}{width=1\textwidth,center=\textwidth}
\begin{tabular}[c]{clllllllll}
 \multicolumn{2}{c}{}  &  \multicolumn{8}{c}{\textit{p}}   \\
\midrule
\textsc{Censoring} & \textsc{Method} \qquad   & $10$ & $20$ & $50$ & $100$ & $200$ & $400$ & $800$ & $1200$ \\ 
   \midrule 
    $\kappa=0.85$ & MF-VB
    & $0.016$ \small{[307]} & $0.046$ \small{[719]} & $0.070$  \small{[712]} & $0.134$  \small{[828]} & $0.104$  \small{[630]} & $0.129$  \small{[675]} & $0.170$  \small{[610]} & $0.219$  \small{[571]} \\
    & PFM-VB & $0.091$  \small{[85]} & $0.237$  \small{[192]} & $0.236$  \small{[155]} & $0.262$  \small{[140]} & $0.061$  \small{[47]}   & $0.039$  \small{[13]}   & $0.062$  \small{[8]}  & $0.100$ \small{[7]} \\
    & EP & $0.014$  \small{[4]} & $0.018$  \small{[4]} & $0.068$  \small{[6]} & $0.270$  \small{[5]}  & $0.175$  \small{[4]}  & $0.252$  \small{[3]}  & $0.327$  \small{[2]} & $0.488$  \small{[2]}  \\
\midrule  
    $\kappa=0.50$ & MF-VB
    & $0.002$ \small{[50]}  & $0.004$ \small{[87]}   & $0.007$ \small{[97]}   & $0.020$ \small{[194]} & $0.034$ \small{[303]} & $0.050$ \small{[236]} & $0.111$ \small{[230]} & $0.197$ \small{[248]} \\
    & PFM-VB & $0.009$  \small{[14]}   & $0.016$  \small{[23]}   & $0.017$  \small{[21]}    & $0.024$ \small{[32]}    & $0.021$  \small{[15]}    & $0.033$ \small{[6]}   & $0.086$  \small{[4]}   & $0.165$  \small{[3]}  \\
    & EP & $0.008$  \small{[4]}   & $0.008$  \small{[3]}    & $0.024$  \small{[4]}    & $0.050$  \small{[4]}  & $0.090$  \small{[5]}    & $0.142$ \small{[4]} & $0.295$  \small{[4]}  & $0.459$ \small{[4]}   \\
\midrule  
    $\kappa=0.15$ & MF-VB
    & $0.001$   \small{[14]}   & $0.001$   \small{[19]}   & $0.001$   \small{[25]}  & $0.003$   \small{[46]}   & $0.019$   \small{[130]}  & $0.043$ \small{[111]}  & $0.126$  \small{[22]}   & $0.253$  \small{[7]}   \\
    & PFM-VB & $0.001$   \small{[5]}   & $0.002$   \small{[6]}  & $0.002$    \small{[7]}  & $0.003$   \small{[8]}   & $0.018$   \small{[7]}   & $0.042$   \small{[3]}  & $0.126$   \small{[3]}   & $0.253$   \small{[3]}   \\
    & EP & $0.002$  \small{[3]}   & $0.002$   \small{[3]}   & $0.003$   \small{[3]}   & $0.005$   \small{[3]}  &  $0.023$    \small{[4]}   & $0.052$   \small{[4]}   & $0.149$   \small{[4]}   & $0.278$   \small{[3]}   \\
\midrule
\end{tabular}
\end{adjustbox}
\end{table*}

Consistently with \citet{chopin_2017} and despite the limited theory guarantees, EP emerges as the most accurate solution in  Figure~\ref{figure:1} since its discrepancy from the Monte Carlo estimates under i.i.d.\ sampling is negligible in all regimes, ranging from $p < n_0 $ to $p \geq n_0$. Nonetheless, as highlighted in Figure~\ref{figure:2}, the per-iteration runtimes of EP are linear in $p$ for high  dimensions, as opposed to the quadratic growth in previously-available implementations, whereas those of MF-VB and PFM-VB are essentially constant and much lower. The impact of such a scaling in the cost per iteration on the overall runtime is relatively moderate in the regimes we focus on, as  illustrated in Table~\ref{table_Niter} which displays the overall runtimes required by each routine to perform not only the total iterations, but also  pre- and post-computations. Nonetheless, the higher per-iteration cost of EP becomes clearly appreciable when $p$ exceeds a few hundreds, as seen in Figure~\ref{figure:2} and Table~\ref{table_Niter}. Therefore, from a computational perspective, the variational routines are more effective and scalable alternatives in high-dimensional settings. This is especially true for  PFM-VB  which, as expected from the theory in \citet{fasano2019}, attains the same accuracy of EP when $p \gtrsim 2 n_0$, but with a lower computational effort. On the contrary, MF-VB  is not competitive with   EP  in terms of accuracy and does not yield remarkable improvements in runtimes relative to  PFM-VB in high dimensions. Notice that, although PFM-VB and MF-VB have the same per-iteration cost, the former does not allow for a joint updating of the approximating densities for the local variables, thus requiring a \texttt{for} cycle in basic \texttt{R} implementations that yields an increment in the per-iteration runtimes under PFM-VB,  due to software-related inefficiencies. This drawback can be seen from Table~\ref{table_Niter}, especially in those $p < n_0$ regimes where PFM-VB requires a moderate number of iterations to reach convergence.

Table~\ref{table_Niter} also displays the number of iterations needed by the three deterministic approximation procedures to reach convergence. Following common practice   \citep[e.g.,][]{blei_2017}, the convergence for the two variational methods is intended as observing a difference below a suitable threshold between the ELBO of two consecutive iterations; see \citet{fasano2019, consonni_2007} and the code in the GitHub repository for an expression of the ELBO under PFM-VB and MF-VB. The results presented in Table~\ref{table_Niter}  correspond to a threshold of $10^{-3}$ on the logarithm of the ELBO. Conversely, as discussed in Section~\ref{sec_4.3.2},  EP routines do not share the same monotonicity and convergence guarantees that characterize variational routines. Therefore, the convergence of EP is commonly assessed  in terms of the maximum absolute difference over the set of low dimensional parameters, characterizing the approximate Gaussian sites, between consecutive iterations  \citep[][]{chopin_2017}, for which we employ the same threshold as above. Consistent with the theoretical results in \citet{fasano2019}, Table~\ref{table_Niter} provides evidence on the fact that the number of iterations needed by PFM-VB to reach convergence of the ELBO decays to one as $p$ grows to infinity, while also displaying the phenomenon reminiscent   of  double-descent noticed in Table~\ref{table_runtimes_MC}, thus motivating further research along this line. Interestingly, the empirical results in Table~\ref{table_Niter} also suggest that the number of iterations required by EP does not grow with $p$. These analyses point toward EP as a default strategy, while suggesting PFM-VB as a feasible accurate alternative in high dimensions.

\section{Discussion and future research directions}\label{sec_6}

This review article provides a novel unified methodological and computational framework for Bayesian inference within a wide class of routinely-used regression models under a similarly broad set of prior distributions, which include the Gaussian one. Such an important gap in the literature is covered by first expressing the likelihoods associated with probit, tobit, multinomial probit and their extensions as special cases of a single formulation, and then generalizing previous findings for specific models in  e.g., \citet{durante_2019,fasano2020,fasano2021closed,cao2020scalable,benavoli2020skew,benavoli2021unified} to prove SUN \citep{ARELLANO_VALLE_2006}  conjugacy for any representation that admits such a  unified likelihood. This yields  general versions of past and more recent computational methods, previously proposed only for some specific members of the general class and with a focus on Gaussian priors.  These include data-augmentation Gibbs samplers, i.i.d.\ sampling schemes, VB approximations and scalable EP implementations. 

Due to the relevance of the models considered, such a review is expected to catalyze increasing interest by applied, computational and methodological researchers, and will hopefully motivate further research advancements along the directions opened by the results in Sections~\ref{sec_2}--\ref{sec_5}. For instance, the closed-form expressions in Section~\ref{sec_3.2}  for inference under the exact SUN posterior provide additional motivations to stimulate ongoing research aimed at developing accurate and fast methods to evaluate cumulative distribution functions of high-dimensional Gaussian distributions. In fact, any advancement along this direction and in sampling from multivariate truncated normals can be directly applied to conduct posterior inference via the closed-form results in Section~\ref{sec_3.2}, for increasingly larger sample sizes $\bar{n}+\bar{n}_0$, beyond small-to-moderate settings. This would be also useful for estimation of possible unknown parameters in the covariance matrices $\bar{\bSigma}_{1}$ and $\bar{\bSigma}_{0}$, via numerical maximization of the marginal likelihood $p(\by)$ in \eqref{eq15}. As clarified in Sections  \ref{sec_2.1}--\ref{sec_2.4} such matrices are often parameterized by a one-dimensional or low-dimensional vector of parameters, and hence, can be effectively estimated via direct maximization of $p(\by)$ when its evaluation is computationally practical. The availability of a closed-form expression  \eqref{eq15} for $p(\by)$ and of i.i.d.\ sampling schemes from $(\bbeta \mid \by)$ as in  \eqref{eq14.add} can be also useful to facilitate full Bayesian inference for  $\bar{\bSigma}_{1}$ and $\bar{\bSigma}_{0}$ when the associated parameters are assigned a prior. For example, leveraging $p(\by)$ it is possible to derive collapsed Metropolis--Hastings schemes to sample from the posteriors of $\bar{\bSigma}_{1}$ and $\bar{\bSigma}_{0}$ after integrating out $\bbeta$ analytically, thereby improving mixing of data-augmentation MCMC \citep[][]{park2009partially}; see also \citet{chan2009mcmc} for effective MCMC methods to infer $\bar{\bSigma}_{0}$ under identifiability constraints. These advancements are beyond the scope of this review, but provide a research direction that is worth further exploration. Finally, it is also interesting to include hyperpriors for the scale parameters of the Gaussian or, more generally, SUN prior, that  yield to scale-mixture representations inducing shrinkage in high dimensions \citep{carvalho2010horseshoe}. Since most of these constructions rely on conditionally Gaussian priors, the results in the present review may be useful to obtain improved theoretical and practical performance in state-of-the-art implementations of the models in Section~\ref{sec_2} under sparse settings; see \citet{onorati2022extension} for recent advancements along these lines, motivated by SUN conjugacy.

Although  \eqref{eq1} already encompasses several models of interest, further generalizations of such a likelihood and of the conjugacy results in Section~\ref{sec_3} can be considered. For instance, it is possible to extend \eqref{eq1} to any version of the models in Sections~\ref{sec_2.1}--\ref{sec_2.4} that arise from censoring or rounding of the Gaussian latent utilities into a generic truncation region. As discussed in the Supplementary Materials, such a mechanism is directly related to the generative construction of the broader class of selection distributions (SLCT)  in \citet{arellano2006unified}. Hence, following the same general reasoning, it seems natural to prove SLCT conjugacy for this broader family of likelihoods. For example, this has been done  in \citet{kowal2021conjugate} and \citet{king2021warped} by extending the ideas in  \citet{durante_2019} and \citet{fasano2021closed}  to static and dynamic rounded-data situations. These generalizations can be considered to prove similar conjugacy results for any extension of \eqref{eq1} which incorporates truncation into a finite region. It would be also interesting to extend the recent conjugacy results under skew-elliptical link functions, such as skew-$t$    \citep{branco2001general,azzalini2003distributions,gupta2003multivariate}, to the proposed general framework. In fact, the unified skew-elliptical distribution \citep{arellano2010multivariate} has a general density expression that shares the product form of the likelihood \eqref{eq1}, with a probability density function multiplied by a cumulative distribution function computed in some appropriate linear transformation of the parameter. This class of distributions admits as particular cases both the SUN and the unified skew-$t$ distribution, and thus also the skew-$t$. Motivated by these results, \citet{zhang2021tractable} proved that unified skew-elliptical distributions  are conjugate to probit and multinomial probit with skew-elliptical link functions, thus suggesting that this result may hold for any regression model based on skew-elliptical utilities. 

As proved in a recent article by \citet{durante_2023}, suitable generalizations of skew-normal distributions  also provide more accurate limiting laws and skew-modal approximations for generic posterior distributions --- beyond those considered in this article --- with an improved convergence rate relative to the one achieved by the classical Bernstein--von Mises theorem based on limiting Gaussians.

Finally, it shall be also emphasized that the class of models discussed  in Section~\ref{sec_2} arguably encompasses one of the broadest set of formulations that appear in econometrics \citep{Greene2003} and social sciences \citep[][]{demaris2004regression}. Nonetheless, routine applications of such models under a Bayesian perspective have lagged behind the growing interest in Bayesian statistics. This review not only  clarifies that the posterior distributions induced by the likelihoods of these models belong to a known class of variables, but also that such conjugacy results hold for a broader set of priors and for various extensions of classical probit, tobit and multinomial probit that are of direct relevance in econometrics and social sciences. This will hopefully boost the  routine-use of these Bayesian models in applied research and motivate the development  of even more flexible versions which still belong to likelihood  \eqref{eq1}, including, for example, random effects and graphical models \citep[e.g.,][]{jones2005}.

\renewcommand{\theequation}{S.\arabic{equation}}
\renewcommand{\thesection}{S\arabic{section}}  
\setcounter{section}{0}
\setcounter{equation}{0}
\section*{SUPPLEMENTARY MATERIALS}
\vspace{-5pt}
The Supplementary Materials contain (i) further examples of relevant regression models whose likelihood can be re-written as in Equation   \eqref{eq1}, (ii)  proofs of Lemma~\ref{lem1} and Theorem~\ref{teo1}, and (iii) a detailed discussion on the computational costs of the routines to derive the approximations in Section~\ref{sec_4.3}. Code can be found at \url{https://github.com/niccoloanceschi/TobitSUN}.

\vspace{-5pt}
\section{Further extensions to skewed link functions, non-linear models and dynamic settings}\label{sec_2.4_sup}
Although the models discussed in Sections~\ref{sec_2.1}, \ref{sec_2.2} and \ref{sec_2.3} cover the most widely-implemented formulations in the literature, as highlighted in Sections~\ref{sec_2.4.0}--\ref{sec_2.4.3} several additional extensions of these representations to skewed, non-linear, dynamic and other contexts can be reframed within the  likelihood in~\eqref{eq1}.

\subsection{Inclusion of skewed latent utilities}\label{sec_2.4.0}

\subsubsection{Skewed extensions of linear regression and multivariate linear regression}
It is possible to include skewness within the formulations in Section~\ref{sec_2.1}, while still remaining in the class of models whose likelihood can be expressed as in \eqref{eq1}. Recalling \citet{sahu2003new} and \citet{azzalini2005skew}, this can be done by assuming that $(y_i \mid \bbeta) \sim \textsc{sn}(\bx^{\intercal}_i\bbeta, \sigma^2, \alpha)$, independently for each $i=1, \ldots, n$, where $\textsc{sn}(\bx^{\intercal}_i\bbeta, \sigma^2, \alpha)$ denotes the skew-normal distribution \citep{azzalini1985} with location $\bx^{\intercal}_i\bbeta$, scale $\sigma^2$ and shape parameter $\alpha$. This choice implies that
\begin{equation}
\begin{split}
 p(\by \mid \bbeta)  &\propto\prod\nolimits_{i=1}^{n} \phi(y_i- \bx^{\intercal}_i\bbeta; \sigma^2) \Phi(\alpha(y_i- \bx^{\intercal}_i\bbeta); \sigma^2)\\
& \ \  = \phi_{n}(\by-\bX\bbeta; \sigma^2\bI_n) \Phi_n(\alpha\by-\alpha\bX\bbeta; \sigma^2\bI_n),
\end{split}
\label{eq4}
\end{equation}
which coincides again with  \eqref{eq1}, when $\bar{n}_1=\bar{n}_0=n$, $\bar{\by}_1=\by$, $\bar{\by}_0=\alpha\by$, $\bar{\bX}_1=\bX$, $\bar{\bX}_0=-\alpha\bX$ and $\bar{\bSigma}_1=\bar{\bSigma}_0=\sigma^2 \bI_n$. Inclusion of skewed responses from more elaborated distributions such as the multivariate skew-normal \citep{AZZALINI_dellaValle_1996,azzalini1999}, the extended multivariate skew-normal \citep{arnold2000,arnold2002}, the closed skew-normal family \citep{gonzalez2004,gupta2004} and the SUN  \citep{ARELLANO_VALLE_2006}, is also possible and yields again special cases of Equation  \eqref{eq1};  see, e.g., \citet{canale_2016}.

\subsubsection{Skewed extensions of probit, multivariate probit and multinomial probit}
The inclusion of skewed link functions is possible also under probit, multivariate probit and multinomial probit models. This direction has been effectively explored by \citet{chen1999new} and \citet{bazan2010framework} with a main focus on basic probit models, and can be again reframed within the general formulation in~\eqref{eq1}. For example, in the context of univariate probit regression, skewness can be incorporated by replacing the Gaussian latent utilities with skew-normal ones; namely $(z_i \mid \bbeta) \sim \textsc{sn}(\bx^{\intercal}_i\bbeta, \sigma^2, \alpha)$, independently for $i=1, \ldots, n$. As a consequence, the binary response data $y_i=\mathbbm{1}(z_i>0)$ are Bernoulli variables with probabilities $\mbox{pr}(y_i=1 \mid \bbeta)=\mbox{pr}(z_i>0 \mid \bbeta) \propto \Phi_2[(\bx^{\intercal}_{i}\bbeta, 0)^{\intercal}; \mbox{diag}(\sigma^2, 1)+\sigma \alpha (1+\alpha^2)^{-1/2}({\bf 1}_2{\bf 1}^{\intercal}_2-\bI_2)]$, whose expression follows directly from  the  cumulative distribution function of the skew-normal; see e.g., \citet{gonzalez2004,ARELLANO_VALLE_2006,azzalini2010prospective,azzalini_2013}, and \citet{arellano2021}. Leveraging the same results, it also  follows that $\mbox{pr}(y_i=0 \mid \bbeta)=\mbox{pr}(z_i<0 \mid \bbeta) \propto \Phi_2[(-\bx^{\intercal}_{i}\bbeta, 0)^{\intercal}; \mbox{diag}(\sigma^2, 1)-\sigma \alpha (1+\alpha^2)^{-1/2}({\bf 1}_2{\bf 1}^{\intercal}_2-\bI_2)]$. Let $\bX_i=[(2y_i-1)\bx_i, {\bf 0}]^{\intercal}$ and $\bSigma_i=\mbox{diag}(\sigma^2, 1)+(2y_i-1)\sigma \alpha (1+\alpha^2)^{-1/2}({\bf 1}_2{\bf 1}^{\intercal}_2-\bI_2)=\bSigma+(2y_i-1)\bLambda$, the joint likelihood for the binary response data can be then expressed as follows
\begin{equation}
\begin{split}
p(\by \mid \bbeta)&\propto \prod\nolimits_{i=1}^{n}\Phi_2(\bX_{i}\bbeta; \bSigma_i)\\
&\ \ \  =\Phi_{2n}(\bX \bbeta;  \bI_n \otimes\bSigma+\mbox{diag}(2\by-{\bf 1}_n)\otimes\bLambda),
\end{split}
\label{eq8}
\end{equation}
with $\bX=(\bX^{\intercal}_1, \ldots, \bX^{\intercal}_n)^{\intercal}$, $\bSigma=\mbox{diag}(\sigma^2, 1)$, and $\bLambda=\sigma \alpha (1+\alpha^2)^{-1/2}({\bf 1}_2{\bf 1}^{\intercal}_2-\bI_2)$. As a consequence, Equation \eqref{eq8} is again a special case of \eqref{eq1}, after setting  $\bar{n}_1=0$, $\bar{n}_0=2n$, $\bar{\by}_0={\bf 0}$, $\bar{\bX}_0=\bX$ and $\bar{\bSigma}_0=\bI_n \otimes\bSigma+\mbox{diag}(2\by-{\bf 1}_n)\otimes\bLambda$. Similar derivations can be considered to incorporate skewness within multivariate and multinomial probit via multivariate skew-normal \citep{AZZALINI_dellaValle_1996},  closed skew-normal \citep{gonzalez2004,gupta2004} or unified skew-normal \citep{ARELLANO_VALLE_2006} latent utilities. Some of these choices have not  yet been explored to induce skewed link functions for multivariate and multinomial extensions of classical probit regression. Nonetheless,  all these variables have cumulative distribution functions proportional to those of multivariate Gaussians, evaluated at a linear combination of $\bbeta$, and, hence, induce likelihoods which can be again expressed as special cases of the general framework in \eqref{eq1}.

\subsubsection{Skewed extensions of tobit regression}
As for the models presented in Sections~\ref{sec_2.1} and \ref{sec_2.2}, also tobit regression admits extensions to skewed contexts. This generalization has been explored, for example, by \citet{hutton2011modelling} who replace the Gaussian assumption for each $(z_i \mid \bbeta) \sim \mbox{N}(\bx^{\intercal}_i\bbeta, \sigma^2)$ with the skew-normal $(z_i \mid \bbeta) \sim \textsc{sn}(\bx^{\intercal}_i\bbeta, \sigma^2, \alpha)$, independently for every  unit $i=1, \ldots, n$. Recalling the derivations for the skewed extensions of the models in Sections~\ref{sec_2.1}--\ref{sec_2.2}, this assumption implies that the contribution to the likelihood for the $i$--th data point is proportional to $(\phi(y_i- \bx^{\intercal}_i\bbeta; \sigma^2) \Phi[\alpha(y_i- \bx^{\intercal}_i\bbeta); \sigma^2])^{\mathbbm{1}(y_i>0)}(\Phi_2[(-\bx^{\intercal}_{i}\bbeta, 0)^{\intercal}; \mbox{diag}(\sigma^2, 1)-\sigma \alpha (1+~\alpha^2)^{-1/2}$ $({\bf 1}_2{\bf 1}^{\intercal}_2-\bI_2)])^{\mathbbm{1}(y_i=0)}$. Therefore, letting $\mbox{diag}(\sigma^2, 1)=\bSigma $ and \smash{$\sigma \alpha (1+\alpha^2)^{-1/2}({\bf 1}_2{\bf 1}^{\intercal}_2-\bI_2)=\bLambda$}, yields
\begin{equation}
\begin{split}
	p(\by \mid \bbeta)&\propto \prod\nolimits_{i: y_i>0} \phi(y_i- \bx^{\intercal}_i\bbeta; \sigma^2) \Phi(\alpha(y_i- \bx^{\intercal}_i\bbeta); \sigma^2)\\
	&\qquad \cdot  \prod\nolimits_{i: y_i=0}  \Phi_2[(-\bx^{\intercal}_{i}\bbeta, 0)^{\intercal}; \bSigma-\bLambda]\\
	&\quad = \phi_{n_1}(\by_1{-}\bX_1\bbeta; \sigma^2\bI_{n_1})\\
	&\qquad \ \cdot \Phi_{n_1+2n_0}(\alpha [\by_1^{\intercal}, {\bf 0}^{\intercal}]^{\intercal}-(\alpha\bX_1^{\intercal},\bX^{\intercal}_{0})^{\intercal} \bbeta; \bSigma_0),
	\end{split}
					\label{eq10}
	\end{equation}
where $n_1$, $n_0$, $\by_1$ and $\bX_1$ are defined as in \eqref{eq9}, whereas $\bX_{0}$ is a $2n_0 \times p$ design matrix obtained by stacking $2 \times p$ row blocks $\bX_i=(\bx_i, {\bf 0})^{\intercal}$ for those units with $y_i=0$, while $\bSigma_0$ is a block-diagonal matrix with blocks $\bSigma_{0[1,1]}= \sigma^2\bI_{n_1}$, $\bSigma_{0[2,2]}= \bI_{n_0} \otimes \bSigma-\bI_{n_0} \otimes \bLambda$. Hence, to express \eqref{eq10} as a particular case of \eqref{eq1} it suffices to set $\bar{n}_1=n_1$, $\bar{n}_0=n_1+2n_0$, $\bar{\by}_1=\by_1$, $\bar{\by}_0=\alpha [\by_1^{\intercal}, {\bf 0}^{\intercal}]^{\intercal}$, $\bar{\bX}_1=\bX_1$, $\bar{\bX}_0=-(\alpha\bX_1^{\intercal},\bX^{\intercal}_{0})^{\intercal} $, $\bar{\bSigma}_1=\sigma^2\bI_{n_1}$ and $\bar{\bSigma}_0=\bSigma_0$. Recalling discussions in Sections~\ref{sec_2.1}--\ref{sec_2.2}, these derivations can be directly applied to incorporate skewness in type {\rm II}--{\rm V} tobit models \citep{amemiya1984}, also under more general distributions which extend the original skew-normal \citep[e.g.,][]{ARELLANO_VALLE_2006}.

\vspace{-5pt}

\subsection{Inclusion of generic thresholds}\label{sec_2.4.1}
All the results presented in Sections~\ref{sec_2.1}--\ref{sec_2.3} hold, under minor changes, when replacing the commonly-used zero threshold  with a generic one $z_{\textsc{t}}$, possibly varying between units. For instance, in probit regression this modification implies that $\mbox{pr}(y_i=1 \mid \bbeta)=\Phi(-z_{\textsc{t}}+\bx^{\intercal}_i\bbeta)$, thus providing the joint likelihood $\prod_{i=1}^n \Phi[(2y_i-1)(-z_{\textsc{t}}+\bx_i^{\intercal}\bbeta)]=\Phi_{n}(-z_{\textsc{t}}(2\by-{\bf 1}_n)+\mbox{diag}(2\by-{\bf 1}_n)\bX\bbeta; \bI_n)$, which coincides with expression  \eqref{eq1} after letting $\bar{n}_1=0$, $\bar{n}_0=n$, $\bar{\by}_0=-z_{\textsc{t}}(2\by-{\bf 1}_n)$, $\bar{\bX}_0=\mbox{diag}(2\by-{\bf 1}_n)\bX$ and $\bar{\bSigma}_0=\bI_n$. Similar derivations apply to multivariate probit, multinomial probit, tobit, and their skewed extensions.

By contrast, all models relying on truncations to a finite interval of the form $[z_{1\textsc{t}}, z_{2\textsc{t}}]$ do not induce likelihoods that can be rewritten as in \eqref{eq1}. Nonetheless, these versions are  less frequent than those presented in Sections~\ref{sec_2.1}--\ref{sec_2.3} and, as discussed in Section~\ref{sec_6}, the SUN conjugacy results presented for the general class of models whose likelihoods admit the form \eqref{eq1}, are useful to motivate similar extensions for a generic truncation mechanism. In fact, as discussed in \citet{arellano2006unified}, the SUN family belongs itself to an even more general class of selection distributions (SLCT) whose construction rely on cumulative distribution functions evaluated at generic  intervals. This result has been recently leveraged by  \citet{kowal2021conjugate} and \citet{king2021warped} to extend the original  SUN conjugacy properties presented by \citet{durante_2019} and \citet{fasano2021closed}  for probit regression and its multivariate dynamic extensions, respectively, to rounded/categorical data  where truncation is  in finite intervals \citep[e.g.,][]{jeliazkov2008fitting}. These modifications can be  extended to prove the SLCT conjugacy for generalizations of 
\eqref{eq1} which admit truncation to any finite interval.

\subsection{Inclusion of non-linear effects}\label{sec_2.4.2}
Another key extension of the models presented in Sections~\ref{sec_2.1}--\ref{sec_2.3} can be obtained by including non-linearities within the predictor. A common solution to accomplish such a goal is to replace  $f(\bx_i)=\bx_i^{\intercal}\bbeta$ with the generic basis expansion $f(\bx_i)=\bg(\bx_i)^{\intercal} \bbeta$, where $\bg(\bx_i)=[g_1(\bx_i), \ldots, g_k(\bx_i)]^{\intercal}$ are pre-specified non-linear basis functions, such as splines \citep[see e.g.,][]{holmes2001bayesian,lang2004bayesian}. Including this extension within the general framework in Equation \eqref{eq1} poses no difficulties since it is sufficient to replicate the derivations for the models presented in Sections~\ref{sec_2.1}, \ref{sec_2.2} and \ref{sec_2.3} with $\bx_i=(x_{i1}, \ldots, x_{ip})^{\intercal}$ replaced by $\tilde{\bx}_i=[g_1(\bx_i), \ldots, g_k(\bx_i)]^{\intercal}$, for each $i=1,\ldots, n$.

Alternatively, one can model directly~$[f(\bx_1), \ldots, f(\bx_n)]^{\intercal} \in \mathbbm{R}^n$ via a Gaussian process \citep[e.g.,][]{Rasmussen2006}. This direction has been commonly explored in the context of the models presented in Sections~\ref{sec_2.1}--\ref{sec_2.3}   \citep[e.g.,][]{kuss2005assessing,de2005bayesian,girolami2006,nickisch2008approximations,riihimki2012nested,cao2020scalable,benavoli2020skew,benavoli2021unified}, and can be also reframed within formulation  \eqref{eq1}. In particular, by assuming, without loss of generality, no overlap in $\bx_1, \ldots, \bx_n$, the Gaussian process construction with mean function $m(\cdot)$ and covariance kernel $K(\cdot, \cdot)$ implies that $[f(\bx_1), \ldots, f(\bx_n)]^{\intercal}$ is jointly distributed as a $\mbox{N}_n(\bxi, \bOmega)$ with $\bxi=[m(\bx_1), \ldots, m(\bx_n)]^{\intercal}$ and covariance matrix $\bOmega$ having entries $\bOmega_{ii'}=K(\bx_i, \bx_{i'})$, for each $i=1, \ldots, n$ and $i'=1, \ldots, n$. This representation can be alternatively rewritten as $\tilde{\bX} \bbeta$, where $\bbeta=[f(\bx_1), \ldots, f(\bx_n)]^{\intercal} \sim \mbox{N}_n(\bxi, \bOmega)$ and $\tilde{\bX}=\bI_{n}$. Therefore, letting $\tilde{\bx}_i$ denote an $n \times 1$ vector with value $1$ in position $i$ and $0$ elsewhere, for each $i=1, \ldots, n$, it is possible to consider Gaussian process extensions of the models in Sections~\ref{sec_2.1}--\ref{sec_2.3}, while still remaining within the general framework in \eqref{eq1}.


\subsection{Inclusion of dynamic structure}\label{sec_2.4.3}
Time--varying extensions of the models  in Sections~\ref{sec_2.1}, \ref{sec_2.2} and \ref{sec_2.3} are common  in the literature  \citep[e.g.,][]{manrique1998simulation,andrieu2002particle,naveau2005skewed,chib2006inference,soyer2013bayesian,fasano2021closed}. These extensions often appear as generalizations of the  original dynamic linear model having observation equation $(\by_t \mid \bbeta_t) \sim \mbox{N}_m(\bX_t \bbeta_t, \bSigma_t)$, independently for every time $t=1, \ldots, n$, and state equations $(\bbeta_t \mid \bbeta_{t-1}) \sim \mbox{N}_p(\bG_t\bbeta_{t-1}, \bW_t)$, independently for any $t=1, \ldots, n$, where $\bX_t$, $\bSigma_t$, $\bG_t$, and $\bW_t$ are known system matrices, whereas  $\bbeta_0 \sim \mbox{N}_p(\ba_0, \bP_0)$. This building-block construction implies that the contribution to the likelihood of $\by_t$, for every  $t=1, \ldots, n$, is $p(\by_t \mid \bbeta)= \phi_m(\by_t- \bX_t\bbeta_t; \bSigma_t)=\phi_m(\by_t- \tilde{\bX}_t\bbeta; \bSigma_t)$, where $\bbeta=(\bbeta^{\intercal}_1, \ldots, \bbeta^{\intercal}_n)^{\intercal}$ and $ \tilde{\bX}_t=\bv^{\intercal}_t \otimes \bX_t$, with $\bv_t$ denoting a $n \times 1$ indicator vector having value $1$ in position $t$ and $0$ elsewhere. Therefore, this representation can be directly interpreted as a particular version of the multivariate linear regression model in Equation \eqref{eq3} with covariance matrix possibly changing across the time units. Such a connection allows to directly recast the joint likelihood $p(\by \mid \bbeta)$ of $\by=(\by^{\intercal}_1, \ldots, \by^{\intercal}_n)^{\intercal}$ within \eqref{eq1}. Clearly, this result holds for any subsequence $\by^{\intercal}_{1:t}=(\by^{\intercal}_1, \ldots, \by^{\intercal}_t)^{\intercal}$, with $t=1, \ldots, n$, thereby facilitating online derivation of the filtering $p(\bbeta_t \mid \by_{1:t})$, predictive $p(\bbeta_{t+1} \mid \by_{1:t})$ and smoothing $p(\bbeta \mid \by)$ distributions via the Gaussian-Gaussian conjugacy  implied by the observation and state equations \citep{kalman1960}.

The above results have been recently extended by \citet{fasano2021closed} for deriving the first analog of the classical Kalman filter \citep{kalman1960} within the context of multivariate dynamic probit models with Gaussian states, leveraging the SUN--probit conjugacy properties proved in \citet{durante_2019}. Recalling \citet{fasano2021closed} and adapting the notation to the one in this article, the contribution to the likelihood of $\by_t$, for every time $t=1, \ldots, n$, can be expressed as $p(\by_t \mid \bbeta)=\Phi_m(\bB_t\bX_t \bbeta_t; \bB_t \bSigma_t \bB_t)$, where $\bX_t$ and $\bB_t$ are defined as in \eqref{eq6}, with $i$ replaced by time $t$, whereas  $\bSigma_t$ is a possibly time-varying covariance matrix among the latent utilities $(z_{t1}, \ldots, z_{tm})^{\intercal}$. Recalling the derivations considered for the Gaussian dynamic setting, the expression for $p(\by_t \mid \bbeta)$ can be alternatively rewritten as $p(\by_t \mid \bbeta)=\Phi_m(\tilde{\bX}_t \bbeta; \bB_t \bSigma_t \bB_t)$, with $\tilde{\bX}_t=\bv^{\intercal}_t \otimes (\bB_t\bX_t)$, which shows again the direct connection between this dynamic formulation and its static counterpart in \eqref{eq6}, thereby allowing to recast the induced joint likelihood for  $\by=(\by^{\intercal}_1, \ldots, \by^{\intercal}_n)^{\intercal}$ and its subsequences $\by^{\intercal}_{1:t}=(\by^{\intercal}_1, \ldots, \by^{\intercal}_t)^{\intercal}$, $t=1, \ldots, n$, within Equation \eqref{eq1}.

These results clearly hold also for the dynamic extensions of models \eqref{eq2}  and  \eqref{eq5}, which represent the univariate versions of \eqref{eq3} and \eqref{eq6}, respectively, thus simply requiring to set $m=1$ in the above derivations. Similarly, multinomial probit \eqref{eq7} and tobit \eqref{eq9} observation equations, along with skewed extensions  (\eqref{eq4}, \eqref{eq8}, \eqref{eq10}), can be again reframed within \eqref{eq1} since all these constructions are characterized by contributions to the likelihood for each time $t=1, \ldots, n$ having the same form of those associated with the statistical units $i=1, \ldots, n$ in the static counterparts of such models presented  in Sections~\ref{sec_2.1}--\ref{sec_2.3}, under suitable specifications of the design and covariance matrices.

As a final remark, it is worth emphasizing that~\eqref{eq1} naturally encompasses any combination of the models discussed in Sections~\ref{sec_2.1}--\ref{sec_2.4}. For example, if  $\by_i=(y_{i1},y_{i2},y_{i3},y_{i4})^{\intercal}$, where $y_{i1}$, $y_{i2}$, $y_{i3}$ and $y_{i4}$ are from models in \eqref{eq2}, \eqref{eq5}, \eqref{eq7} and \eqref{eq9}, respectively, for each $i=1, \ldots, n$, then, leveraging the derivations in Sections~\ref{sec_2.1}--\ref{sec_2.3}, it directly follows that the joint likelihood for the vector $\by=(\by^{\intercal}_1, \ldots, \by^{\intercal}_n)^{\intercal}$ still belongs to \eqref{eq1}.

\section{Proofs of Lemma 1 and Theorem 1}\label{sec_3_p}

\begin{proof}[Proof of Lemma~\ref{lem1}] To prove Lemma~\ref{lem1}, first notice that, by Bayes rule 
$$p(\bbeta \mid \bar{\by}_1) \propto p(\bbeta) p(\bar{\by}_1 \mid \bbeta),$$
where $p(\bar{\by}_1 \mid \bbeta)= \phi_{\bar{n}_1}(\bar{\by}_1-\bar{\bX}_1 \bbeta;\bar{\bSigma}_1)$, whereas $p(\bbeta)$ is the SUN density in \eqref{eq11}. Leveraging Gaussian-Gaussian conjugacy, it follows that the product between $ \phi_{\bar{n}_1}(\bar{\by}_1-\bar{\bX}_1 \bbeta;\bar{\bSigma}_1)$ and the density term $\phi_{\bar{p}}(\bbeta - \bxi;\bOmega) $ in \eqref{eq11} is proportional to $\phi_{\bar{p}}[\bbeta-(\bOmega^{-1}+\bar{\bX}^{\intercal}_1 \bar{\bSigma}{}^{-1}_1\bar{\bX}_1)^{-1}(\bOmega^{-1} \bxi+\bar{\bX}^{\intercal}_1 \bar{\bSigma}{}^{-1}_1 \bar{\by}_1);(\bOmega^{-1}+\bar{\bX}^{\intercal}_1 \bar{\bSigma}{}^{-1}_1\bar{\bX}_1)^{-1}]=\phi_{\bar{p}}(\bbeta-\bxi_1; \bOmega_1)$, where 
\begin{equation*}
\begin{split}
\bxi_1&=(\bOmega^{-1}+\bar{\bX}^{\intercal}_1 \bar{\bSigma}{}^{-1}_1\bar{\bX}_1)^{-1}(\bOmega^{-1} \bxi+\bar{\bX}^{\intercal}_1 \bar{\bSigma}{}^{-1}_1 \bar{\by}_1),\\
\bOmega_1&=(\bOmega^{-1}+\bar{\bX}^{\intercal}_1 \bar{\bSigma}{}^{-1}_1\bar{\bX}_1)^{-1}=\bomega_1 \bar{\bOmega}_1 \bomega_1,
\end{split}
\end{equation*}
Therefore, $p(\bbeta \mid \bar{\by}_1) $ is proportional to the product between this updated Gaussian density and the cumulative distribution function term $ \Phi_{\bar{n}}(\bgamma + \bDelta^\intercal \bar{\bOmega}{}^{-1} \bomega^{-1}(\bbeta-\bxi); \bGamma - \bDelta^\intercal \bar{\bOmega}{}^{-1} \bDelta)$ of the SUN density in \eqref{eq11}, which can be also re-expressed as $ \Phi_{\bar{n}}[\bs^{-1}_1\bgamma + \bs^{-1}_1\bDelta^\intercal \bar{\bOmega}{}^{-1} \bomega^{-1}(\bbeta-\bxi); \bs^{-1}_1(\bGamma - \bDelta^\intercal \bar{\bOmega}{}^{-1} \bDelta)\bs^{-1}_1]$, where $\bs^{-1}_1$ is defined as in Lemma~\ref{lem1}. To prove that this product yields to the SUN  kernel in Lemma~\ref{lem1}, rewrite $\bs^{-1}_1\bgamma + \bs^{-1}_1\bDelta^\intercal \bar{\bOmega}{}^{-1} \bomega^{-1}(\bbeta-\bxi)$ as $\bs^{-1}_1\bgamma- \bs^{-1}_1\bDelta^\intercal \bar{\bOmega}{}^{-1} \bomega^{-1}\bxi+\bs^{-1}_1\bDelta^\intercal \bar{\bOmega}{}^{-1} \bomega^{-1}\bbeta$, and then sum and subtract $\bs^{-1}_1\bDelta^\intercal \bar{\bOmega}{}^{-1} \bomega^{-1}\bxi_1$ inside this expression to obtain 
\begin{equation*}
\begin{split}
&\bs^{-1}_1[\bgamma+\bDelta^\intercal \bar{\bOmega}{}^{-1} \bomega^{-1}(\bxi_1-\bxi)]+\bs^{-1}_1\bDelta^\intercal \bar{\bOmega}{}^{-1} \bomega^{-1}(\bbeta-\bxi_1)\\
&=\bgamma_1+\bs^{-1}_1\bDelta^\intercal \bar{\bOmega}{}^{-1} \bomega^{-1}(\bbeta-\bxi_1)\\
&=\bgamma_1+\bs^{-1}_1\bDelta^\intercal \bar{\bOmega}{}^{-1} \bomega^{-1} \bomega_1\bar{\bOmega}_1\bar{\bOmega}{}^{-1}_1 \bomega{}^{-1}_1(\bbeta-\bxi_1)\\
&=\bgamma_1+\bDelta_1^\intercal \bar{\bOmega}_1{}^{-1} \bomega_1^{-1}(\bbeta-\bxi_1),
\end{split}
\end{equation*}
with $\bgamma_1$ and $\bDelta_1$ defined as $\bgamma_1=\bs^{-1}_1[\bgamma+\bDelta^\intercal \bar{\bOmega}{}^{-1} \bomega^{-1}(\bxi_1-\bxi)]$ and $\bDelta_1= \bar{\bOmega}_1\bomega_1\bomega^{-1} \bar{\bOmega}{}^{-1}\bDelta\bs^{-1}_1$, respectively. In order to conclude the proof, note that the correlation matrix within the cumulative distribution function term can be also rewritten as 
\begin{equation*}
\begin{split}
&\bs^{-1}_1(\bGamma - \bDelta^\intercal \bar{\bOmega}{}^{-1} \bDelta)\bs^{-1}_1\\
&=\bs^{-1}_1(\bGamma-\bDelta^\intercal \bar{\bOmega}{}^{-1} \bDelta)\bs^{-1}_1+\bDelta_1^\intercal \bar{\bOmega}{}^{-1}_1 \bDelta_1-\bDelta_1^\intercal \bar{\bOmega}{}^{-1}_1 \bDelta_1\\
&=\bs^{-1}_1[\bGamma+\bDelta^\intercal (\bar{\bOmega}{}^{-1}\bomega^{-1} \bOmega_1\bomega^{-1}\bar{\bOmega}{}^{-1}-\bar{\bOmega}{}^{-1})\bDelta]\bs^{-1}_1\\
&\qquad \qquad \qquad \qquad \qquad \qquad \qquad \qquad \qquad  \quad -\bDelta_1^\intercal \bar{\bOmega}{}^{-1}_1 \bDelta_1,
\end{split}
\end{equation*}
which corresponds to $\bGamma_1-\bDelta_1^\intercal \bar{\bOmega}{}^{-1}_1 \bDelta_1$, with $\bGamma_1$ defined as in Lemma~\ref{lem1}. This proves that the kernel $p(\bbeta) p(\bar{\by}_1 \mid \bbeta)$ of the posterior coincides with that of a SUN having parameters $\bxi_1$, $\bOmega_1$, $\bDelta_1$, $\bgamma_1$ and $\bGamma_1$ specified as in  Lemma~\ref{lem1}.
\end{proof}

\begin{proof}[Proof of Theorem~\ref{teo1}]
The proof of Theorem~\ref{teo1} simply requires to combine  Lemma~\ref{lem1}  with an adaptation of Corollary 4 in \citet{durante_2019}. In particular, by direct application of the Bayes rule, it follows that 
$$p(\bbeta \mid {\by}) \propto p(\bbeta) p(\by  \mid \bbeta) \propto  [p(\bbeta)p(\bar{\by}_1 \mid \bbeta)]p(\bar{\by}_0  \mid \bbeta).$$ 
Therefore, the posterior distribution $p(\bbeta \mid {\by})$ can be obtained by first updating the SUN prior $p(\bbeta)$ with $p(\bar{\by}_1 \mid \bbeta)\propto\phi_{\bar{n}_1}(\bar{\by}_1-\bar{\bX}_1 \bbeta;\bar{\bSigma}_1)$, and then use such a conditional density $p(\bbeta \mid \bar{\by}_1)$ as an intermediate prior to be updated with the likelihood $p(\bar{\by}_0 \mid \bbeta)\propto\Phi_{\bar{n}_0}(\bar{\by}_0+\bar{\bX}_0 \bbeta; \bar{\bSigma}_0)$ of $\bar{\by}_0$ for obtaining the final posterior. By direct application of Lemma~\ref{lem1}, it follows that  $p(\bbeta \mid \bar{\by}_1)$ is the density of the  $\mbox{SUN}_{\bar{p}, \bar{n}}(\bxi_1,\bOmega_1,\bDelta_1,\bgamma_1,\bGamma_1)$ with parameters defined as in Lemma~\ref{lem1}. Therefore, to conclude the proof, it is sufficient to prove that the updating of this intermediate prior with the likelihood $p(\bar{\by}_0 \mid \bbeta)\propto\Phi_{\bar{n}_0}(\bar{\by}_0+\bar{\bX}_0 \bbeta; \bar{\bSigma}_0)$ for $\bar{\by}_0$ yields again to a $\mbox{SUN}_{\bar{p}, \bar{n}+\bar{n}_0}(\bxi_{\pst},\bOmega_{\pst},\bDelta_{\pst},\bgamma_{\pst},\bGamma_{\pst})$ with $\bxi_{\pst}$, $\bOmega_{\pst}$, $\bDelta_{\pst}$, $\bgamma_{\pst}$ and $\bGamma_{\pst}$ defined as in Theorem~\ref{teo1}. This result follows directly from an adaptation of Corollary 4 in \citet{durante_2019}; refer also to Theorem~1  in  \citet{fasano2020}.  In particular,  replacing $\bD$ with $\bar{\bX}_0$ and $\bI_n$ with $\bar{\bSigma}_0$ in Corollary 4 by \citet{durante_2019}, under a $\mbox{SUN}_{\bar{p}, \bar{n}}(\bxi_1,\bOmega_1,\bDelta_1,\bgamma_1,\bGamma_1)$ prior, yields to the expressions for $\bxi_{\pst}$, $\bOmega_{\pst}$, $\bDelta_{\pst}$ and $\bGamma_{\pst}$ in Theorem~\ref{teo1}. Inclusion of the offset $\bar{\by}_0$ in the proof of Corollary 4 by \citet{durante_2019} poses no difficulties since it directly enters the SUN truncation parameter, thereby providing the expression for $\bgamma_{\pst}$ in Theorem~\ref{teo1}.
\end{proof}

\section{Computational costs}\label{sec_4.3.3}
To further extend the analysis of the approximate schemes  in Sections~\ref{sec_4.3.1}--\ref{sec_4.3.2}, we discuss the associated cost per-iteration focusing, for the ease of notation, on the classical probit regression $\prod_{i=1}^{n}\Phi(\bx^{\intercal}_i\bbeta)^{\mathbbm{1}(y_i=1)}[1-\Phi(\bx^{\intercal}_i\bbeta)]^{\mathbbm{1}(y_i=0)}$ as in \eqref{eq5}, with a spherical Gaussian prior $p(\bbeta)=\phi_{p}(\bbeta; \omega^2 \bI_{p})$. Note that, as discussed in Section~\ref{sec_2.2}, $n=\bar{n}_0$ and $p=\bar{p}$ when such a model is written as a special case of likelihood \eqref{eq1}. Besides providing one of the most widely implemented formulations within the class of models whose likelihood can be expressed as in \eqref{eq1}, this choice is also motivated by the fact that detailed costs per-iteration of effective MF-VB and PFM-VB implementations have been already derived in \citet{fasano2019} under probit regression with a Gaussian prior. Moreover, it gives the opportunity to show that currently-reported per-iteration costs of EP for the same class of models and priors \citep{chopin_2017} can be further reduced, thus making also EP more scalable to high dimensions.

For deriving the costs of MF-VB, PFM-VB and EP it shall be emphasized that, in probit regression, such approximations rely on $c=1, \ldots, n$ and, therefore, $n_c=\bar{n}_c=1$ for any $c$. Under EP   \citep{chopin_2017} this choice is a common practice which follows from the factorized form of the likelihood, whereas in MF-VB \citep{consonni_2007} it is implied by the formulation of the optimization problem and is a direct consequence of the conditional independence among the unit-specific latent utilities. Instead, for PFM-VB \citep{fasano2019} such a setting is not enforced. Nonetheless, it provides a convenient specification which is in line with MF-VB and EP solutions, and also facilitates posterior inference by only requiring to deal with univariate truncated normals. 

Under the aforementioned settings, Appendix A of \citet{fasano2019} provides a detailed discussion of the per-iteration cost for both  MF-VB and also PFM-VB, which is  $\mathcal{O}(n \cdot \mbox{min}\{n,p\})=\mathcal{O}(\bar{n}_0\cdot \mbox{min}\{\bar{n}_0,\bar{p}\})$, after suitable matrix precomputations before running the CAVI routine. Since the mean and variance of univariate truncated normals can be accurately computed at $\mathcal{O}(1)$ cost under standard algorithms \citep[e.g.,][]{Botev_2016}, the most intensive computations in the CAVI routines for MF-VB and PFM-VB correspond to the matrix multiplication operations. These steps can be efficiently implemented by exploiting recursive formulas when updating each univariate truncated normal approximating density conditioned the most recent estimate of the others, in PFM-VB, or of the $\bbeta$ parameters, in MF-VB, thereby leading to an overall per-iteration cost that is either linear or sublinear in $p=\bar{p}$.

As for EP, the currently reported per-iteration cost in probit regression with spherical Gaussian priors is $\mathcal{O}(np^2)=\mathcal{O}(\bar{n}_0\bar{p}^2)$ \citep{chopin_2017}, after suitable precomputations as in  MF-VB and PFM-VB. Intuitively, this increased complexity is due to the fact that, unlike for MF-VB and PFM-VB, not only the expectations but also the $p \times p$ covariance matrices must be updated  and then inverted at every site $c$, for each $c=1, \ldots, n$. Although the specific form of such matrices allows to reduce the common cubic cost into a quadratic one via the application of the Woodbury's formula to avoid direct matrix inversion \citep{chopin_2017}, an $\mathcal{O}(np^2)=\mathcal{O}(\bar{n}_0\bar{p}^2)$ cost can be still computationally impractical in the large $p=\bar{p}$ setting. In fact, as mentioned in the final discussion by \citet{chopin_2017}, even the state-of-the-art implementations of EP are often computationally challenging when $p$ exceeds one thousand. This point is also confirmed in the empirical studies of  \citet{fasano2019}, where the EP implementation within the \texttt{R} package \texttt{EPGLM} by  \citet{chopin_2017} requires more than six hours to reach convergence in a high-dimensional Alzheimer's application with $p=\bar{p}=9036$ and $n=\bar{n}_0=300$. Notably, as a further contribution of the present article, it shall be emphasized that a more scalable EP implementation with per-iteration cost $\mathcal{O}(np \cdot \mbox{min}\{n,p\})=\mathcal{O}(\bar{n}_0\bar{p}\cdot \mbox{min}\{\bar{n}_0,\bar{p}\})$ can be actually derived by leveraging similar results considered in  \citet{fasano2019} for obtaining efficient implementations of  MF-VB and PFM-VB. In particular, this new EP implementation exploits the fact that, under the same reformulation via Woodbury's identity of \citet{chopin_2017}, the site updates do not necessarily require direct computation of the  aforementioned $p \times p$ matrices, since such quantities enter via the inner product with the $p \times n$ design matrices $\bar{\bX}^{\intercal}_{0}$. Hence, when $p=\bar{p}$ is large, it is more convenient to update this product directly, without  storing, updating or multiplying any $p \times p$ matrix. This yields to an $\mathcal{O}(np)=\mathcal{O}(\bar{n}_0\bar{p})$ cost for each site, and to an overall cost for the $n=\bar{n}_0$ site updates of  $\mathcal{O}(n^2p)=\mathcal{O}(\bar{n}^2_0\bar{p})$. In high dimensional settings, when $p \gg n$, this linear cost in $p=\bar{p}$ yields to massive computational gains relative to the original $\mathcal{O}(np^2)=\mathcal{O}(\bar{n}_0\bar{p}^2)$ cost of \texttt{EPGLM}  in \citet{chopin_2017}. For example, by considering the proposed more scalable implementation within the high-dimensional  Alzheimer's application yields to an overall runtime of less than one minute, which is orders of magnitude lower than \texttt{EPGLM} \citep{chopin_2017} that requires, instead, more than six hours. When, instead, $n \gg p$, the linear cost in $n=\bar{n}_0$ of the \texttt{EPGLM} ensures effective implementations. Combining these two scenarios yields to an overall  per-iteration cost of $\mathcal{O}(np \cdot \mbox{min}\{n,p\})=\mathcal{O}(\bar{n}_0\bar{p}\cdot \mbox{min}\{\bar{n}_0,\bar{p}\})$, which is linear in the higher between $n=\bar{n}_0$ and $p=\bar{p}$. To the best of our knowledge, this is the first implementation of EP available in the literature to achieve such a computational efficiency. Despite this, standard EP remains more computationally demanding than  MF-VB and PFM-VB since, as discussed above, also the $p \times p$ matrices need to be updated at each step of the EP routine, either directly or implicitly via the product with $\bar{\bX}^{\intercal}_{0}$.

The above reasoning can be directly applied in order to highlight a similar dependence on sample size and number of predictors in the per-iteration cost of effective MF-VB, PFM-VB and EP implementations under the whole class of models and priors in Sections~\ref{sec_2} and \ref{sec_3} --- as long as $n_c$ and $\bar{n}_c$ are sufficiently small to allow the calculations of the moments for the associated multivariate truncated normals at a negligible cost compared to the one of the matrix operations. This result is illustrated in empirical studies in Section~\ref{sec_5}, with a focus on tobit regression.

\begingroup
\fontsize{11pt}{12pt}\selectfont

\endgroup

\end{multicols}

\end{document}